\documentclass[12pt,preprint]{aastex}

\slugcomment{Accepted for publication in the Astrophysical Journal}

\shorttitle{The Intrinsically X-ray Weak Quasar PHL 1811. I. }
\shortauthors{Leighly et al.}

\begin{document}


\title{The Intrinsically X-ray Weak Quasar PHL 1811. I. X-ray
  Observations and Spectral Energy Distribution\footnote{Based on
  observations obtained with {\it XMM-Newton}, an ESA science mission
  with instruments and contributions directly funded by ESA Member
  States and NASA}}


\author{Karen M.\ Leighly}
\affil{Homer L.\ Dodge Department of Physics and Astronomy, The
  University of   Oklahoma, 440 W.\ Brooks St., Norman, OK 73019}
\email{leighly@nhn.ou.edu}
\author{Jules P. Halpern}
\affil{Department of Astronomy, Columbia University, 550 W.\ 120th
  St., New York, NY 10027-6601}
\author{Edward B.\ Jenkins}
\affil{Princeton University Observatory, Princeton, NJ 08544-1001}
\author{Dirk Grupe}
\affil{Department of Astronomy and Astrophysics, Pennsylvania State
  University, 525 Davey Lab, University Park, PA 16802}
\author{Jiehae Choi\altaffilmark{1} and Kimberly B.\ Prescott}
\affil{Homer L.\ Dodge Department of Physics and Astronomy, The University of
  Oklahoma, 440 W.\ Brooks St., Norman, OK 73019}
\altaffiltext{1}{Current Address: Department of Astronomy,
 New Mexico State University, P.\ O.\ Box 30001, MSC 4500, Las Cruces,
 NM 88003-8001}




\begin{abstract}

This is the first of two papers reporting observations and analysis of
the unusually bright ($m_b=14.4$), luminous ($M_B=-25.5$), nearby
($z=0.192$) narrow-line quasar PHL~1811, focusing on the X-ray
properties and the spectral energy distribution.  Two {\it Chandra}
observations reveal a weak X-ray source with a steep spectrum.
Variability by a factor of 4 between the two observations separated by
12 days suggest that the X-rays are not scattered emission.  The {\it
XMM-Newton} spectra are modelled in the 0.3--5 keV band by a steep
power law with $\Gamma = 2.3\pm 0.1$, and the upper limit on intrinsic
absorption is $8.7 \times 10^{20} \rm cm^{-2}$. The spectral slopes
are consistent with power law indices commonly observed in NLS1s, and
it appears that we observe the central engine X-rays directly.
Including two recent {\it Swift} ToO snapshots, a factor of $\sim 5$
variability was observed among the five X-ray observations reported
here.  In contrast, the UV photometry obtained by the {\it XMM-Newton}
OM and {\it Swift} UVOT, and the {\it HST} spectrum reveal no
significant UV variability.  The $\alpha_{ox}$ inferred from the {\it
Chandra} and contemporaneous {\it HST} spectrum is $-2.3 \pm 0.1$,
significantly steeper than observed from other quasars of the same
optical luminosity.  The steep, canonical X-ray spectra, lack of
absorption, and significant X-ray variability lead us to conclude that
PHL 1811 is intrinsically X-ray weak.  We also discuss an accretion
disk model, and the host galaxy of PHL~1811.

\end{abstract}




\section{Introduction}

The standard model for Active Galactic Nuclei (AGN) and Quasi-Stellar
Objects (QSOs) proposes that the broad-band optical and UV continuum
originates in an accretion disk.  The X-ray emission is a separate
component, produced in a corona located in the vicinity of the disk,
that creates the X-rays by inverse-Compton-scattering the disk
photons.  This broad-band continuum is then thought to illuminate the
gas that forms the broad-line region, causing it to emit lines via
photoionization.

Despite the commonality of the origin of the optical through X-ray
continuum emission and emission lines, there are theoretical reasons
that the spectra should vary among 
individual objects.  The origin of differences may be extrinsic; for
example, the brightness and, to some extent, the shape of the
continuum spectrum of the accretion disk should vary with viewing
angle \citep[e.g.,][]{ln89}.  The origin may also be intrinsic.  Even
for very simple disk models, in which the spectrum is constructed from
a sum of annuli locally emitting as black bodies, the continuum
spectrum from the accretion disk should depend on the black hole mass
and accretion rate \citep[e.g.,][]{fkr92}.  More sophisticated accretion
disk models that include a range of physical processes expected to be
important also predict a range of shapes \citep[e.g.,][]{kfm98}.  Another
complication is that the type of accretion disk present is predicted
to depend on the  accretion rate relative to the Eddington value
\citep[e.g.,][]{chen95}, and the type of accretion disk can be different
at different radii \citep[e.g.,][]{sz94}.  The X-ray emitting corona
adds another dimension of complication, as its origin and geometry are
not very well understood.  Thus, in principle, the coronal emission
may be important or not  depending on how much of the accretion energy
is funneled to it, and indeed, we see evidence for a range of coronal 
activity in X-ray novae \citep[e.g.,][]{kb04}.

In addition to the theoretical expectation of a range of predicted
spectral energy distributions among AGN, there is observational
evidence that such a range exists.  In a study of the multiwavelength
properties of an X-ray-selected heterogeneous sample of quasars,
\citet{elvis94} observed a wide range of spectral energy distributions
(SEDs).  \citet{wilkes94}, and more recently 
\citet{bechtold03}, \citet{strateva05}, and \citet{steffen06}, found
that $\alpha_{ox}$, the point-to-point slope between the 2500\AA\/ and
2 keV, is inversely correlated with the UV luminosity.  While some of
the large range of spectral energy distributions can be accounted for
by extrinsic effects such as reddening and absorption
\citep[e.g.,][]{blw00}, that certainly cannot account for the behavior
of all objects. 

PHL~1811 is a nearby ($z=0.192$), luminous ($M_B=-25.5$) narrow-line
quasar. PHL~1811 was first cataloged as a blue object in the
Palomar-Haro-Luyten plate survey \citep{hl62}.  It then rediscovered
in the optical followup of the VLA Faint Images 
of the Radio Sky at Twenty Centimeters (FIRST) survey
\citep{white97,bwh95}.  It is extremely bright (B=14.4, R=14.1); it is
the second brightest quasar at $z>0.1$ after 3C~273.  Being so bright, it is a
very good background source for studies of the intergalactic and
interstellar medium; furthermore, a {\it FUSE} observation found its
spectrum to have a rare Lyman-limit system that has been studied by
\citet{jenkins03} and \citet{jenkins05}.  It was odd, however, that
such a bright quasar was not detected in the ROSAT All Sky Survey
(RASS). In comparison with other quasars of its luminosity, the
expected RASS count rate is about $0.5 \rm\, s^{-1}$; we
placed an upper limit of $1.3 \times 10^{-2} \rm \, counts\, s^{-1}$
\citep{lhhbi01}.  A pointed {\it BeppoSAX} observation detected the
object, but it was still anomalously weak. Too few photons were
obtained in the {\it BeppoSAX} observation to unambiguously determine
the cause of the X-ray weakness; \citet{lhhbi01} speculated that
either it is intrinsically X-ray weak, or it is a nearby
broad-absorption line quasar and the X-ray emission is absorbed, or
it is highly variable, and we caught it both times in a low state.  

In this paper and the companion paper \citep[][hereafter Paper
 II]{leighly06} we report the results of several UV and X-ray
  observations of PHL~1811 designed to explore the origin and
 consequences  of the X-ray
 weakness of this object.  First, coordinated {\it Chandra} and {\it
 HST} observations were made in 2001.  In 2004, an {\it XMM-Newton}
 observation was made, and most recently, PHL~1811 was the target of
 two {\it Swift} target-of-opportunity observations.  In \S 2 we
 describe the results and analysis of the {\it Chandra}, {\it
 XMM-Newton}, and {\it Swift}  observations, as well as the results of
 a three-day optical  photometry run at MDM Observatory. We also
 compare the {\it XMM-Newton} spectrum of PHL~1811 with those from
 other NLS1s.  In \S 3 we  comment on the long time-scale X-ray and UV
 variability.  We  present an updated spectral energy 
distribution in \S 4.  In \S 5, we discuss the nature of the intrinsic
 X-ray weakness and present an accretion disk model for PHL~1811.  We
also comment on the apparent spiral host galaxy discovered in the
 image presented by \citet{jenkins05}.  We summarize our findings in
 \S 6.  Paper~II describes the {\it HST} and ground-based  optical and
 UV observations, and presents {\it Cloudy} models that 
 explore the unusual emission-line properties.   Some
 of the results  were presented  previously in  \citet{lhj04},
 \citet{clm05}, and  \citet{prescott06}.    We assume a flat Universe
 with $H_0=70\rm\, km  \,s^{-1}\,Mpc^{-1}$ and $\Omega_{vac}=0.73$
 unless otherwise specified. 

\section{Observations and Analysis}

\subsection{Chandra Observations and Analysis}

The {\it Chandra} observations were made in imaging spectroscopy mode
with the image of PHL~1811 placed on the ACIS-S3 detector.  The
observing log is given in Table 1.  We verify
that the position of the X-ray source is consistent with that of the
quasar (Fig.\ 1).

The level 2 events files were recreated using the standard
procedure. The small correction for the time-dependent gain was
applied using the corr$\_$tgain program, and the correction for the
time-dependent ancillary matrix was made using the IDL program
acisabs.pro\footnote{http://www.astro.psu.edu/users/chartas/xcontdir/xcont.html}.
The total count rates observed within a circular region 3.94 arc
seconds in radius  were $9.7 \times 10^{-3} \rm \, counts\, s^{-1}$
and $4.0 \times 10^{-2} \rm \, counts\, s^{-1}$ from the first and
second observations, respectively.  These rates are low enough that
pile-up is negligible.  Between 0.3 and 9.8 keV, a total of 81 and 374
photons were obtained.  Based on the background collected from
source-free areas of the chips, we expect 1 and 3 of these photons to 
originate in the background.  Thus we can conclude that we observe a
significant change in flux in the object by about a factor of four 
between the two observations separated by 12 days.

A sufficient number of photons were collected in the second
observation to look for short time scale variability.  We  used
the Bayesian Blocks program available in the ISIS
software\footnote{http://space.mit.edu/CXC/analysis/SITAR/index.html},
but did not find any indication of variability during the observation
at significance levels greater than about 1 sigma.  We also tried
binning the light curve with 100-second bins, and then grouping
together by hand points that appeared low or high.  The $\chi^2$ for a
constant model was 13.5 for 6 degrees of freedom, indicating that
variability was marginally detected with confidence of 96.4\%, although
the probability that this variability is significant may be lower
because the results are biased by preselecting the bin sizes.  Note
that the $\Delta\chi^2=6.63$ uncertainty on the fitted constant model is
14\%, indicating that we are only sensitive to variations larger than
this value at the 99\% confidence level.  Such high amplitude
variations are rare but not unprecedented in luminous NLS1s
\citep[e.g.,][]{leighly99a}.  

The spectra were accumulated and grouped so there were $\sim 20$
photons in each bin.  We first fitted each spectrum between 0.3 and 5
keV separately with a power law and fixed Galactic column of $3.76
\times 10^{20}\rm \, cm^{-2}$, obtained using the HEASARC $n_H$
tool\footnote{http://heasarc.gsfc.nasa.gov/cgi-bin/Tools/w3nh/w3nh.pl}.
We note that the Galactic Ly$\alpha$ line in the medium-resolution UV
spectrum of PHL~1811 \citep{jenkins05} is consistent with this value
of $N_H$.  This model fits both spectra well, yielding photon indices
of $2.01^{+0.37}_{-0.36}$ and $2.58^{+0.19}_{-0.18}$ for the first and
second {\it Chandra} observations, respectively.  These photon indices
are consistent with those observed from NLS1s by {\it ASCA}
\citep{leighly99b}.  Note that the {\it ASCA} photon indices were
taken from models spanning the $\sim 0.5$--10 keV band that include a
soft excess, warm absorber, and iron line as necessary.  Thus, the {\it
Chandra} spectra from PHL~1811, fit over 0.3--5.0 keV (0.36--5.96 rest
frame), are consistent with the hard X-ray power law found in {\it
  ASCA} spectra of NLS1s. 

The best-fitting photon index is steeper for the second, brighter
observation, suggesting that shape of the spectrum has changed between
the two observations.  However, the uncertainties indicate that the
difference is not statistically significant. Fitting the spectra
simultaneously and using the F-test shows that the improvement in the
fit represented by a change in the photon index is significant at only
the 68\% confidence level. Fig.\ 2 shows these model fits.

In order to see if there is any evidence for intrinsic absorption, we
next fit the spectrum from the second observation with a model
consisting of a power law, absorption fixed at the Galactic value, and
absorption in the rest frame of the quasar.  We find no improvement in
the fit and the best-fitting value of additional intrinsic absorption
is zero. The $\Delta \chi^2=2.71$ upper limit on the intrinsic
absorption column is $1.8 \times 10^{20}\rm \, cm^{-2}$. The upper 
limit is rather low despite the poor photon statistics in the spectrum
because the spectrum is convex (Fig.\ 2). 

The convex residuals of the power law fit to the second observation
suggest the  presence of a soft excess (Fig.\ 2).  Soft
excess components are common in the spectra of NLS1s, and in the case
of poor statististics can be fit adequately by a black body model
\citep[e.g.,][]{leighly99b}.  We add a black body component to the
power law model for the second spectrum and find that the fit improves
by $\Delta\chi^2=4.9$, and the residuals are flat (Table 2).  However,
the improvement in the fit is not statistically significant; the F
test shows that the improvement in fit is significant at only the 71\%
confidence level. Thus, we cannot conclude that a soft excess is
present because of the poor statistics. 

The power law index for the power law plus black body fit to the
second observation is flatter than for the power law alone ($2.22 \pm
0.34$ versus $2.58^{+0.19}_{-0.18}$), and is now completely consistent
with that of the first observation ($2.01^{+0.37}_{-0.36}$).  This
suggests that the spectral variability originates as an emergence of
the black body component when the object is brighter.  To investigate
this possibility, we fit both spectra simultaneously with a power law
plus blackbody model, fixing the normalization of the black body for
the first observation to zero.  The model fits the data adequately
($\chi^2_\nu=0.67$ for 16 degrees of freedom (d.o.f.)).  The
jointly-fit photon index  is $2.12\pm 0.25$, and for this
parameterization, the normalizations of the power law differ by a
factor of 3.5.  

Spectral ratios provide a complementary and model-independent approach
to the question of spectral variability.  We rebin the spectrum from
the second observation to the binning of the first spectrum and take
their ratio (Fig.\ 2 inset).  This shows that the spectral variability is
predominately in the softest band, supporting our hypothesis that a
variable soft excess is responsible for the spectral variability.
In this case, the variability of the power law component is a
factor of 3.5.  The $\chi^2$ for a constant ratio model is 3.8 for 3
degrees of freedom which significant at the 71\% confidence level.

To summarize the results of the {\it Chandra} observations, we find
conclusive evidence for factor of $\sim 4$ variability between the two
observations separated by 12 days.   Detailed analysis is hampered by
poor statistics; however, we find marginal evidence ($2 \sigma$) for
variability on times scales of thousands of seconds in the second
observation when the object was brighter.  Spectral fitting reveals a
steep spectrum with no evidence for intrinsic absorption, and 
marginal evidence for spectral variability between the observations,
with the spectrum becoming steeper when the object is brighter.   The
spectral fitting and the ratio of the spectra suggest that the
spectral variability is caused by the emergence of a soft excess
component when the object is bright.  The measured spectral indices
range between 2.0 and 2.6, depending on the model. They are
consistent with the photon indices measured in {\it ASCA}
observations of NLS1s \citep{leighly99b}.  This fact suggests that we
see the intrinsic X-ray emission from the central engine, and that the
X-rays are powered by inverse Compton scattering of soft photons as in
other AGN.

\subsection{{\it XMM-Newton} Observation and Analysis}

PHL~1811 was observed by {\it XMM-Newton} 1 November 2004 using the
EPIC PN \citep{stru01} and MOS \citep{turner01} instruments and the
optical monitor \citep{mason01}.  The EPIC observations were carried
out using the Thin filter in PrimeFullWindow mode.  The data were
reduced using standard selection criteria.   The object was observed
for 32.1~ks using the MOS detectors and for 27.5~ks using the PN.  The
details of the  observation are given in Table 1.

Background flares are a concern in {\it XMM-Newton} data analysis, and
we observed the background to vary during the observation.  For most of the
observation, the background relatively low and stable.  However, even
at the lowest rate, it appears to be slightly elevated compared with
the quiescent rate\footnote{XMM-Newton User's Handbook \S 3.3.8
http://heasarc.gsfc.nasa.gov/docs/xmm/uhb/node38.html} by a factor of
approximately 2.3 for the PN and 1.4 in the MOS1+MOS2 in both the soft
(0.5--2.0 keV) and hard (2.0--10 keV) bands.  In addition, for the
first 3000 seconds in the PN detector, and the first 5000 seconds for
the MOS detector, there occurred a small background flare that was
higher than the quiescent rate by a factor of  2--3.  We
extracted and analyzed spectra with and without this flaring period,
and conclude that the flare is so small that it does not adversely
affect the results.  Therefore, we analyze the entire exposure.

We extract light curves from the PN and MOS detectors.  The target is
relatively weak, so we use a source extraction region with a radius
corresponding to an encircled energy function of 80\%; the radius was
$27^{\prime\prime}$ for the PN and $23^{\prime\prime}$ for the MOS1
and MOS2.  Background light curves and spectra were extracted from
nearby, source-free regions of the detector.  The background spectrum
is flatter than the source spectrum, and we find that the background
contribution to the emission in the extraction aperture is equal to
that of the target at $\sim 5\rm \, keV$.  The background dominates at
higher energies; therefore,  we  extract light curves in the 0.5--5
keV band.  We use a bin size of 1000 
seconds, which yields an average of $\sim 30$ source counts per bin,
and the  background contributes about 12\% of the total counts. The
resulting background-subtracted light curve is shown in Fig.\ 3.  The
mean count rate is $2.8 \pm 0.17 \times 10^{-3} \rm \, counts\,
s^{-1}$. The light curve is statistically consistent with no
variability ($\chi^2=33.4$ for 30 d.o.f. for a constant model).   Note
that the $\Delta\chi^2=6.63$ uncertainty on the fitted constant model
is 9\%, indicating that we are only sensitive to variations larger than
this value at the 99\% confidence level.  Such high amplitude
variations are rare but not unprecedented in luminous NLS1s
\citep[e.g.,][]{leighly99a}.  

The spectra were extracted using the regions described above, and
grouped so that there were 15 photons per energy bin.  We fit
the PN in the 0.3--5 keV range and MOS  in the 0.5--5 keV band,
simultaneously.  The spectra yield 611, 191 and 204 source photons for
the PN, MOS1 and MOS2.   The spectra are fit very well with a model
consisting of a constant, a power law and Galactic absorption column
described in \S 2.1. The results are listed in Table 3, and the model
fit is shown in Fig.\ 4.   Note that unlike the second {\it Chandra}
observation, the residuals are flat, and there is no evidence for a
soft excess component.

The spectra are adequately fit using a steep power law
($\Gamma=2.3 \pm 0.1$), typical of that from NLS1s observed by {\it
  ASCA} \citep{leighly99b}.  There is no evidence for additional
absorption.  We add a neutral absorption component at the redshift of
PHL~1811 to the model but find no significant reduction in
$\chi^2$.  The 90\% confidence upper limit ($\Delta\chi^2=2.7$) on
additional absorption is $N_H=8.7 \times 10^{20}\rm \, cm^{-2}$.  

The constant in the model is fixed to a value of 1 for the PN
spectrum, and allowed to be free for the MOS1 and MOS2 spectra.  It
can be seen in Table 3 that the best-fitting normalizations of MOS1
and MOS2 spectra are 25\% and 32\% higher than that of the PN.  This
difference cannot be explained by residual calibration uncertainty
between the  {\it XMM-Newton} EPIC instruments as that is now quite low
\citep{xmm52}.  We analyze the X-ray spectra from
another object in the field (located at RA=21h54m41s, Dec=$-$9d26h49m) 
that is about 50\% brighter than PHL~1811.  The normalizations of the
EPIC spectra for this object were  completely consistent with one
another (MOS1 constant: $1.00 \pm 0.12$; MOS2 constant:
$1.00^{+0.13}_{-0.12}$).  Another object in the field of view was
independently analyzed by another of the authors (D.\ Grupe) with the
same result.

While the small numbers of photons in the PHL~1811 spectra mean that
the difference in normalizations among the models for the different
EPIC spectra is not statistically significant, there still seems to be
a problem with the spectra that needs to be understood.  We believe
that the problem originates in optical loading.  The nominal limit for
optical loading using the thin filter is $V\approx 12$ for both the PN
and the MOS \citep{smith04, altieri03}.  PHL 1811 is a fainter optical
source (B=14.4, R=14.1); however, it has a very blue spectrum and is a
very weak X-ray source, and thus is an unusual object compared with
the stellar calibration sources used to determine the loading limits.
Another point that supports our contention that optical loading is
important is the fact the optical loading in the MOS2 is observed to
be about 7\% larger than that in the MOS1, possibly due to variations
in the filter transmission \citep{altieri03}; we also observe a higher
normalization for the MOS2 spectrum.  However, the degree of
contamination by optical photons cannot be very large because the MOS
spectra do not show any observable distortion; a power law fit to them
alone yields an identical photon index as that obtained from the PN.
Nevertheless, given the fact that the MOS spectra are possibly
contaminated by optical loading, we measure the flux from the PN
spectrum, noting that we cannot be sure that this spectrum is
uncontaminated by optical loading as well, and may represent an upper
limit on the flux for this observation.

PHL 1811 was observed using the Optical Monitor with the UVM2 filter.
Ten exposures, each with duration of 2580 seconds, were made.  The SAS
task {\it omichain} was run to reprocess the data.  The count rate
information was extracted from the {\it omichain} output.  In
addition, the count rates were extracted from the images using the
IRAF task {\tt phot} following the procedure described on
http://xmm.vilspa.esa.es/sas/documentation/watchout/uvflux.shtml.  The
count rates were then converted to flux using the conversion factor
for the UVM2 filter ($2.17 \times 10^{-14}\rm \, erg\, s^{-1}\,
cm^{-2}\,$\AA\/$^{\rm -1} \, \rm count^{-1}$).  The results from the
two extraction procedures were consistent to
within 2\%, and therefore hereafter we discuss the results from the
{\it omichain} output.

One of the goals of the OM observation was to look for UV variability.  A
constant model fit to the light curve yielded $\chi^2=16.86$ for 9
degrees of freedom.  Thus, a constant model is rejected at the 94.9\%
confidence level.  However, we do not consider this evidence for
marginally significant UV variability because the fluctuations are
different for the {\it omichain} and IRAF-reduced data.  The mean flux
of the 10 observations was $2.773 \pm 0.007 \times 10^{-14} \rm \,
erg\, s^{-1}\, cm^{-2}\, $\AA\/$^{-1}$.

\subsection{Joint {\it Chandra} and {\it XMM-Newton} Modelling}

In principle, better constraints on spectral variability can be
obtained by jointly fitting the {\it Chandra} and {\it XMM-Newton}
spectra.  We first fit with a power law plus Galactic absorption
model, allowing the normalizations to be free
among the two {\it Chandra} spectra and the {\it XMM-Newton} PN
spectrum, but with the photon indices constrained to be equal.  The
spectra are fit adequately; the $\chi^2$ is 70.5 for 70
degrees of freedom.  The best fitting photon index is
$2.36^{+0.12}_{-0.11}$, again typical of an NLS1 \citep{leighly99b}.
We notice that the normalization for the spectrum from the first {\it
  Chandra} observation is identical to that of the {\it XMM-Newton} PN
spectrum; if we tie these two parameters together, there is no change
in $\chi^2$ ($\chi^2=70.5$ for 71 degrees of freedom).  These two
spectra seem to describe the object in the same state (see also \S 3),
so we leave their parameters tied together throughout the remainder of
this section.    

Next, we allow the photon index of the second, higher flux, {\it
Chandra} observation to be  fit independent of that of the first {\it
  Chandra} spectrum and the {\it XMM-Newton} spectrum.    We obtain a
somewhat better fit with 
$\chi^2=64.1$ for 70 degrees of freedom.  However, the decrease in
$\chi^2$ is significant at only the 63\% level according to the F
test.  The resulting photon indices are $2.22 \pm 0.14$ for the low
state (first {\it Chandra} observation and {\it XMM-Newton} PN
spectrum), and $2.57^{+0.19}_{-0.18}$ for the second, brighter {\it
 Chandra} observation.  These indices are still within the range of
power law indices observed from NLS1s \citep{leighly99b}.  In
addition, it is often found the NLS1 spectra soften as when they are
brighter (e.g., Fig.\ 4 of \citet{leighly99a}); thus, intrinsic
softening of the spectrum is a plausible explanation for the marginal
spectral variability we observe.

It is possible that the variability results from
variable cold absorption. We test this scenario by constraining the
photon indices and normalizations to be the same for all three
spectra, and including a neutral absorption column in the quasar rest
frame in the model, allowing the absorption column to vary.    This
model gives a very poor fit; the reduced $\chi^2=2.57$ for 70 degrees
of freedom. Thus, we can reject on statistical grounds the idea that
the spectral and flux variability originates solely in variable
neutral absorption.  Allowing the normalizations to also be free
yields a good fit ($\chi^2=65.3$ for 69 degrees of freedom), although
it is not a significant improvement over the  no-absorption  model
($\Delta \chi^2=5.2$, significant at the 58\% confidence level
according to the F test).  The additional absorption is consistent
with zero for the brighter {\it   Chandra} observation with an upper
limit of $1.4 \times 10^{20}\rm \, cm^{-2}$, and is equal to
$5.0^{+3.9}_{-3.6} \times 10^{20}\rm \, cm^{-2}$ for the fainter {\it
  Chandra} observation and the {\it   XMM-Newton}
observation. Although this model (variable cold absorption plus
variable power-law normalization) fits the spectra well, the scenario
seems unlikely because it requires that absorber variability
be coordinated with intrinsic flux variability.   

\subsection{The X-ray Spectra of NLS1s; How does PHL 1811 Compare?}  

{\it ASCA} observations of NLS1s demonstrate that their spectra can 
generally be described by a hard power law with average index of $2.19
\pm 0.10$ and a soft excess that can be modeled using a blackbody
component.  The strength of the soft excess varies from object to
object, with the objects having the overall steepest spectra also
showing the highest amplitude variability \citep{leighly99b}. 

{\it XMM-Newton}, with its large effective area, has revolutionized
our understanding of the X-ray spectra of AGNs.  Soft excesses are now
seen to be relatively common in quasars in general
\citep[e.g.,][]{porquet04}.   NLS1 spectra are still found to have
soft excess that sometimes can  be modelled by a dual Comptonization
model, in which soft photons are scattered by Comptonizing media of
two different temperatures.   Examples of
objects with this type of spectrum are Ton~S180 \citep{vaughan02} and
Mrk~896 \citep{page03}. In other cases, the X-ray spectrum is very
complex, with a very prominent soft excess and complex absorption
features at high energies.  These spectra can be modelled using
partial covering \citep[e.g.][]{tanaka04} or reflection \citep{mf04}.
Examples of this type of object include 1H~0707$-$495 \citep{gallo04b,
tanaka04}, IRAS~13224$-$3809 \citep{boller03}, and PHL~1092
\citep{gallo04}.  \citet{gallo06} split NLS1s into two classes:
those with and without significant complex features in their high 
energy spectra.  He finds that objects with complex high energy
spectra tend to be X-ray weak, and proposes that the X-ray weakness is
consistent with either attenuation in the partial covering scenario,
or the focusing X-rays away from our line of sight in the reflection
scenario.

How does the {\it XMM-Newton} spectrum of PHL~1811 compare with those
from other NLS1s?  It is important to note that the quality of our
spectrum is much poorer than those from many {\it XMM-Newton}
observations of NLS1s, such as those mentioned above, due to its low
X-ray flux.  The PN spectrum has 857 photons between 0.3 and 5 keV in
the observed frame, corresponding to 0.36--5.96~keV in the rest
frame. Of these, 246 are background, leaving 611 net source photons.
This spectrum is adequately modelled with a power law and absorption
originating in our Galaxy.  The reduced $\chi^2$ is 0.94 for 51
degrees of freedom, and the photon index is $2.25 \pm 0.15$. The ratio
of the data to the power law plus Galactic absorption model is shown in
Fig.\ 5.  This figure shows that there are no residuals that suggest a
more complex model, and indeed, because the reduced $\chi^2$ is less
than one, a more complex model would over-parameterize the data and
would not be justified statistically.  It is important to note that
this photon index is consistent with the average hard X-ray photon
index observed in {\it ASCA} spectra from NLS1s \citep[$2.19 \pm
0.10$;][]{leighly99b}.  Thus, PHL~1811 resembles an average NLS1
without a soft excess.  The power law X-ray spectrum is believed to be
produced by Compton upscattering of soft photons in a hot plasma. This
is the same process believed to operate in AGN in general; the reason
that the photon index is steeper in NLS1s than in broad-line quasars
because the hot plasma has been Compton cooled
\citep[e.g.,][]{pounds95}.  

Is the lack of complexity in the PHL 1811 spectrum due to the low
flux and poor statistics, or does it really have a simple spectrum?
We can obtain some answers to this question by extracting sufficiently
short segments of {\it XMM-Newton} data from other NLS1s so that we obtain
spectra with approximately 611 photons between 0.3 and 5 keV.  We
perform this exercise on two NLS1s: Ton S180 and 1H~0707$-$495.

Ton~S180 has a spectrum with a mild soft excess.  \citet{vaughan02}
model it using a dual Comptonization model, and it was classified by
\citet{gallo06} as an NLS1 with a ``simple'' spectrum.   
The Ton S180 data are seen in the middle panel of Fig.\ 5.   This
object is so bright that we need an exposure of just 51.52 seconds to
obtain 572 photons between 0.34--5.61 keV (the observed-frame range
corresponding to the rest-frame range of 0.36--5.96 keV for this
$z=0.062$ object).    We fit the spectrum with a power law
plus Galactic absorption, and obtain a good fit, with a photon 
index of $3.04^{+0.17}_{-0.16}$ and $\chi^2_\nu=0.99$ for 33 degrees
of freedom.  The fact that the reduced $\chi^2$ is less than one  means 
that there is no statistical evidence for spectral complexity; in
addition, we see no suggestive residuals in the data-to-model ratio.
However, in contrast to PHL~1811, the photon index is 
significantly steeper than the average from NLS1s observed by {\it
  ASCA}, clearly because in this spectrum we are fitting the soft
excess component predominantly. If this were the only X-ray spectrum
that we had of this object, we would suspect that spectral complexity 
may be present and a hard tail might be seen in a longer observation,
as it is \citep{vaughan02}.  

1H~0707$-$495 has a complex spectrum that was modeled by
\citet{gallo04b} using partial covering, and was classified by
\citet{gallo06} as an NLS1 with a ``complex'' spectrum. Note that this
is the class that \citet{gallo06} observe to be somewhat X-ray weak.
The 1H~0707$-$495 data are seen in the lower panel of Fig.\ 5.  In 
this case, a segment 289 seconds in length yielded a spectrum with
602 photons between 0.34 and 5.73 keV (the observed-frame range
corresponding to the rest-frame range of 0.36--5.96 keV for this
$z=0.041$ object).  We fit the spectrum with a power law plus Galactic
absorption.  In this case, the resulting photon index is even steeper
than for Ton~S180 ($\Gamma=4.17^{+0.18}_{-0.17}$), the reduced
$\chi^2$ is significantly greater than 1 ($\chi^2_\nu=1.61$ for 34
degrees of freedom), and significant residuals are seen in the
ratio. Clearly, despite the poor statistics in the spectrum, the need
for a complex model is evident. It is also clear that although this
object is X-ray weak, the X-ray spectrum is very dissimilar to that of
PHL~1811. 

This exercise shows that the X-ray spectrum of PHL~1811 is clearly 
different than those of two other NLS1s, Ton~S180 and 1H~0707$-$495.
These objects are representative of the complexity of spectra from
NLS1s observed by observed by {\it XMM-Newton}.  The {\it  XMM-Newton}
spectrum from PHL~1811, a simple power law with a photon index
consistent with the  mean hard X-ray index from NLS1s observed by {\it
  ASCA}, suggests that the X-ray emission mechanism is simple 
Compton-upscattering of soft photons in a hot plasma, as in other AGN,
and that the spectrum is unaltered by any extrinsic effects such as partial
covering or reflection. 

It may also be important that PHL~1811 is more UV-luminous than other
NLS1s.  The monochromatic luminosity at 2500\AA\/ is 30.9;  (Paper~II);
thus it is seen to be about 5 times more luminous than the most
luminous object shown in Fig.\ 2 of \citet{gallo06}.  Also, note that
a very similar figure is shown in \citet{leighly01} and \citet{mlk04}.

\subsection{{\it Swift} Observations and Analysis}

PHL 1811 was observed by the {\it Swift} Gamma-Ray Burst Explorer Mission
\citep{gehrels04} on 2005 October 22 starting at 09:52 UT for a total
of 2.5 ks.  The observations were performed simultaneously with the
X-ray Telescope \citep[XRT; ][]{burrows04} in the 0.3-10.0 keV energy
range and the UV-Optical Telescope \citep[UVOT; ][]{roming04} in the
1700-6500\AA~wavelength range.  It was observed again on 2006 May 12
for 1.6 ks.  The details of the observations are given in Table 1.

The XRT data reduction was performed by the task {\it xrtpipeline}
version 0.9.9, which is included in the HEAsoft package 6.0.4. Source
photons were selected in a circle with a radius of 23.4$^{''}$ and the
background photons in a source-free region close by with a radius of 
95$^{''}$. Those photons were extracted and read into separate event
files with {\it XSELECT} version 2.3.
Twenty-two photons were detected during the first observation, and ten
were detected during the second observation for count rates of $8.8
\pm 1.9 \times 10^{-3}\rm \, s^{-1}$ and $6.2 \pm 2.0 \times
10^{-3}\rm\ \, s^{-1}$, respectively.   

In the 2005 October 22 observation, the UVOT photometry was performed
with the  UV filters UVW1, UVM2, and UVW2.  During the 2006 May 12
observation, all six (optical and UV) filters were used.  The details
are given in Table 1.  After the aspect correction the exposures in
each filter were coadded into one image with {\it  uvotimsum} and the
magnitudes and fluxes in each filter were determine with the task {\it
uvotsource}.   The results (observed fluxes, uncorrected for Galactic
reddening)  are listed in Table 4.  

\subsection{MDM Optical Photometry}

Optical photometry data were taken on PHL~1811 at MDM Observatory
using the 1.3 McGraw-Hill telescope on the nights of 2004 October 14,
15 and 16 as part of a project to search for optical variability in
the narrow-line quasar  PHL~1092 \citep{gallo04}.  We used the
thinned CCD ``Templeton'' and the telescope in the f/7.6
configuration, yielding a angular size of
$0.50^{\prime\prime} \rm /pixel$.  The weather was good on October 14, with
typical seeing of $1.7^{\prime\prime}$, although the sky was not
photometric.  The seeing was worse on October 15 (average of
$2.3^{\prime\prime}$) and it was intermittently cloudy.  On October 16, the
weather had deteriorated further, and few usable frames were
obtained.  

We observed PHL~1811 using the I, V, B and U filters, obtaining
several exposures in each filter.  The total number of frames analyzed
were 9, 19, 12 and 11 for the I, V, B and U filters,
respectively. Within each night, the observations were all obtained
within a time span of about 30 minutes, so we could test the interday
optical variability principally between October 14 and October 15, as
the sampling and data quality was much worse on October 16.

The images were reduced using standard IRAF procedures.  Aperture
photometry was performed on PHL~1811 and 4 field stars using an
aperture twice the size of the image PSF FWHM.  The aperture was
chosen to ensure that essentially all of the photons from the object
were measured, especially in cases where the image was slightly
trailed due to clouds and loss of tracking.  The ratio between
PHL~1811 and the field stars was computed and errors were propagated.
PHL~1811 is a bright object, so the mean signal-to-noise ratios in
these ratios are high, ranging from 50 to 210.  

We tested for variability in the ratio light curves using the ``excess
variance'', a technique commonly used in X-ray astronomy.  The excess
variance is a measure of the observed variance in the light curve
subtracting the variance due to measurement errors.  In nearly all
cases, this measure was negative.  We conclude that no evidence for
optical variability was found.

\section{Long Time-scale X-ray and UV Variability}

PHL~1811 has now been observed in the X-ray bandpass seven times
between 1990 and 2006: during the {\it ROSAT} All Sky Survey, by {\it
  BeppoSAX} in 2000 \citep{lhhbi01}, and in the two {\it Chandra}
observation, the {\it XMM-Newton} observation, and the two {\it Swift}
observations reported here.  In Fig.\ 6 we show the long-term light
curve composed of the secure measurements, i.e., the last five
observations.     We do not plot the RASS upper limit
or the {\it BeppoSAX} observation that, with the large detector PSF,
was certainly contaminated by X-ray emission from neighboring objects.
For the {\it Chandra} and {\it XMM-Newton} observations, the X-ray
flux densities were estimated from the best-fitting power-law models,
and the uncertainties were obtained by propagation of the errors on
the normalizations and photon indices.  For the {\it Swift}
observations, the flux densities were estimated from the count rates
using PIMMS\footnote{http://cxc.harvard.edu/toolkit/pimms.jsp}, 
assuming the {\it XMM-Newton} photon index ($\Gamma=2.3$) and Galactic
absorption, and the uncertainties in the flux densities were assumed
proportional to the uncertainty in the count rate.  

We find that PHL~1811 has varied significantly by a factor of $\sim 5$
in this time period.  The first {\it Chandra} observation and the {\it
XMM-Newton} observation found it to be in a relatively low state, with
$\nu F_\nu$ at 2~keV rest frame equal to $\sim 1.0 \times 10^{-14} \rm
\, erg\, s^{-1}\, cm^{-2}$; the fluxes in these two observations are
consistent with one another.  The second {\it Chandra} observation and
the two {\it Swift} observations show a significantly higher flux by a
factor of 3.5--5.5, and the fluxes of these three are roughly
consistent with one another, although the uncertainties are evaluated
differently.  Although there are only five points, these data suggest
that the X-ray flux oscillates between two states that differ
by a factor of 4--5.  The NLS1 1H~0707$-$495 seems to behave the same
way,  oscillating between two flux states that differ by a factor of
$\sim 10$ \citep{leighly02}

As will be discussed in \S 4, quasars with PHL~1811's optical
luminosity are statistically expected have values of
$\alpha_{ox}$\footnote{$\alpha_{ox}$ is defined as the point-to-point
slope between 2500\AA\/ and 2~keV; i.e.,
$\alpha_{ox}=\log(F_{2500}/F_{2keV})/2.61$.}  equal to $-1.6$.  The
dashed line in Fig.\ 6 shows the predicted X-ray flux for
$\alpha_{ox}=-1.6$ based on the UV flux of the {\it HST} spectrum.
Although PHL~1811 varies significantly, it never approachs the nominal
X-ray flux for an object of its UV luminosity.

We have four epochs of UV observations: the {\it HST} STIS
spectroscopic observation made 2001 December 3 (discussed in detail in
Paper II), the {\it XMM-Newton}
OM observation made 2004 November 1, and the two {\it Swift} UVOT
observations made  2005 October 22 and 2006 May 12.   We search for
possible UV variability by comparing the latter three photometry
measurements with the {\it HST} spectrum.

The effective wavelength of the {\it XMM-Newton} OM UVM2 filter is
2310\AA\/.  We plot the observed fluxes on the observed-frame merged
{\it HST} and optical spectrum from Paper II in Fig.\ 7.  We find that
the photometry flux is completely consistent with the observed
spectrum.  

The effective wavelengths of the {\it Swift} UVOT photometry points
are listed in Table 1, and the inferred fluxes are plotted on Fig.\ 7.
Note that the filter transmission functions are different in the {\it
  Swift} UVOT compared with the {\it XMM-Newton} OM, even though the
filter names are the same; hence the
effective wavelengths are somewhat different.    Like the {\it
  XMM-Newton} OM photometry, the 
correspondence between the UVOT photometry and the {\it HST} spectrum
is very good with the exception of the photometry using the UVW2
filter. That filter is especially difficult to calibrate as the 
complete filter transmission curve is not
known\footnote{http://swift.gsfc.nasa.gov/docs/heasarc/caldb/swift/docs/uvot/uvot\_caldb\_filtertransmission\_02.pdf}, 
and because of the paucity of suitable calibration targets. 
Therefore, we do not consider the disagreement with the {\it HST}
spectrum indicative of a spectral change.  We conclude that we observe
no evidence for any UV variability between the four UV observations
that span a time period of 4.5 years.

To summarize, PHL~1811 has now been observed seven times in the X-rays
over a period spanning 16 years.  During this time, the X-ray flux has
been observed to vary by a factor of $\sim 5$, but it 
remains well below that of a typical quasar of its UV luminosity.
In contrast, the four UV observations made over a period of 4.5 years
do not show any convincing evidence of UV variability.

\section{The Spectral Energy Distribution}

We presented the first spectral energy distribution of PHL~1811 in
\citet{lhhbi01}, based on an optical spectrum, a {\it ROSAT} upper
limit, a {\it BeppoSAX} observation, and multiwavelength photometry.
In Fig.\ 8, we present an updated spectral energy distribution.   For
the optical and UV, we used the merged spectrum described in Paper II.  
The {\it Chandra} results are represented by regions on the SED plot
that were constructed using the third joint model fit (Table 2).    The
contours were constructed by successively setting each variable
parameter to its $\Delta \chi^2=2.71$ value and computing the model,
then determining the maximum and minimum of all the models. 

From the best fit {\it Chandra} model and the merged {\it HST} and
optical spectrum, we compute $\alpha_{ox}$ to be $-2.40$ for the first
observation, which was nearest in time to the {\it HST} observation,
and $-2.19$ for the second observation.  As seen in Figs.\ 6 and 7,
the {\it Chandra} and {\it HST} fluxes are comparable with the {\it
  XMM-Newton} and {\it Swift} fluxes, so we confine our discussion to
the {\it Chandra} and {\it HST} data here without loss of generality.  

\citet{wilkes94} compute a regression between optical luminosity and
$\alpha_{ox}$ for a heterogeneous sample of quasars.  Using their
cosmology ($H_0=50 \rm \, km\, s^{-1}\, Mpc^{-1}$, $q_0=0$), we obtain
a luminosity distance for PHL~1811 of 1261~Mpc, and a corresponding
luminosity density at 2500\AA\/ of $1.44 \times 10^{31}\rm \, erg\,
 s^{-1} \, Hz^{-1}$.  Then, using their regression, we
predict $\alpha_{ox}$ to be $-1.6$.  We plot the predicted X-ray flux
assuming this value of $\alpha_{ox}$ on Fig.\ 8, as well as a vertical
bar that indicates the range of X-ray luminosities observed by
\citet{wilkes94}. 

More recently, \citet{strateva05} and \citet{steffen06} updated the
$\alpha_{ox}$ regression 
using a large sample of optically-selected active galaxies that span a
large range in redshift and luminosity, yet have highly complete X-ray
data.  For their cosmology ($H_0=70 \rm \, km\, s^{-1}\, Mpc^{-1}$,
$\Omega_M=0.3$, and $\Omega_{\Lambda}=0.7$) we obtain a luminosity
distance of $936 \rm \,Mpc$, and a corresponding luminosity density at
at 2500\AA\/ of $7.92 \times 10^{30}\rm \, erg\, s^{-1} \,
Hz^{-1}$. Their regression yields a predicted $\alpha_{ox}$ of $-1.60$
also.  From their Fig.\ 4, we find that the envelope of $\alpha_{ox}$
observed for quasars of this luminosity spans approximately $-1.75$ to
$-1.4$.  Our observed X-ray luminosity densities are factors of
130--450 below the high value, and 13--45 below the low value.  Thus,
PHL~1811 is observed to be significantly X-ray weak compared with
other quasars. 

\citet{blw00} compile the distribution of
$\alpha_{ox}$\footnote{\citet{blw00} use an alternative definition of
$\alpha_{ox}$ using the flux density at 3000\AA\/ rather than
2500\AA\/.  They estimate that $\alpha_{ox}(2500)=1.03
\alpha_{ox}(3000) - 0.03\alpha_{u}$, where $\alpha_u$ is the slope of
the spectrum between 2500 and 3000\AA\/.} from the PG quasar sample
studied by \citet{bg92}.  They find a suggestion of a bimodal
distribution with 10 of the 87 objects classified as X-ray weak with
$\alpha_{ox} \leq -2$.  Then, they find there to be a connection between
$\alpha_{ox}$ and the presence of significant \ion{C}{4} absorption
lines, such that most of the the soft X-ray weak objects have
absorption-line equivalent widths greater than 5\AA\/.  They infer
these results to imply that X-ray absorption is the primary origin of
soft X-ray weakness in AGN.

Clearly, PHL 1811 does not follow the trend observed by \citet{blw00}.
It is soft X-ray weak, but, as discussed in Paper II, there is no
evidence for any significant intrinsic \ion{C}{4} absorption lines.  In
the \citet{blw00} sample, objects with similar $\alpha_{ox}$ as
observed from PHL~1811 have a \ion{C}{4} absorption line equivalent
widths of 5--20 \AA\/. 

Furthermore, the X-ray spectrum shows no evidence for intrinsic
absorption.  If low-column-density absorption were present, we expect
to observe a flat spectrum; in contrast, for a single power law
model, we measure photon indices of 2--2.6, consistent with unabsorbed
quasars, and an upper limit on intrinsic absorption of $8.7 \times
10^{20}\rm\, cm^{-2}$.  

Another possibility is that a high-column-density or
Compton-thick absorber is present in our line of sight, so that we see
no direct continuum emission.  Our spectra do not probe high enough
energies to see whether there is a highly absorbed component, as has
been found in BALQSOs \citep[e.g.,][]{gall02,green01}.  If the
continuum emission were completely absorbed, then the X-ray spectrum
that we see might have been scattered into our line of sight (i.e.,
similar to a Seyfert 2 galaxy or BALQSO).  Electron scattering 
is energy independent, so in this scenario we would expect to see the
intrinsic power law with attenuated flux.  In Seyfert 2s,
the electron scattering occurs over an extended area, and thus while
the intrinsic X-ray emission may vary, the scattered emission does not
because variability is washed out as it scatters over the
extended region.  The fact that  we see significant variability
between the two {\it Chandra} observations, separated by 12 days,
argues that we are not seeing scattered light.  This conclusion
depends on the compactness of the electron-scattering mirror; if
unusually compact, observation of variability would be possible.  
Regardless, the observation of significant variability between two
observations separated by twelve days from this luminous quasar
suggests that we are seeing the intrinsic emission from the AGN.

Narrow-line Seyfert 1 galaxies are known for their high-amplitude
X-ray variability \citep[e.g.,][]{leighly99a}.  As discussed in
\citet{lhhbi01}, it was possible, at that time, when we had only
the RASS upper limit and the {\it BeppoSAX} data, that we had
coincidentally only observed PHL~1811 while it was in a transient low
state.  In this paper, we report five more X-ray observations, all of
which find it to be a significantly weak X-ray source.  Thus, the
probability that we coincidentally observe it in a low state is
decreasing.  PHL~1811 appears to be intrinsically X-ray weak.

\section{Discussion}

\subsection{PHL 1811 is Intrinsically X-ray Weak}

In \citet{lhhbi01}, we reported the first X-ray detection of
PHL~1811 by {\it BeppoSAX}.  That observation showed that PHL~1811
appeared to be X-ray weak, but the 65-net-photon spectrum was not
sufficient to determine the origin of the X-ray weakness.  We
presented three alternatives for the X-ray weakness: 1.\ PHL~1811 is a
BALQSO, and the X-ray emission is absorbed; 2.\ since PHL~1811 is an
NLS1, it is highly X-ray variable, and we happened to catch it in a
low state; 3.\ PHL~1811 is intrinsically X-ray weak.  The {\it HST}
observation discussed in Paper II and {\it Chandra} observations
reported here show that it is not a BALQSO, as there is no evidence
for UV absorption lines.  Furthermore, there is no evidence for
absorption in the X-ray spectrum, and the significant variability
between the two {\it Chandra} observations suggests that the X-ray
emission is not scattered.  So the first hypothesis is firmly ruled
out. 

We can never conclusively rule out the second hypothesis, that
PHL~1811 is highly X-ray variable and we always just happen to catch
it in a low state.  However, it has now been observed seven times between
$\sim 1990$ (during the {\it ROSAT} All Sky Survey) and 2006 May (in a
{\it Swift} observation), and it has never been observed to be bright.  In
fact, since it has already varied by a factor of $\sim 5$ among the
observations reported here, it may have already been as bright as it can
get.   It seems increasingly unlikely that it will ever be as X-ray
bright as other quasars. 

Thus, we conclude PHL~1811 is intrinsically X-ray faint.  Most quasars
are bright X-ray sources, so what property of the PHL~1811 central
engine causes it to be X-ray faint?  As we pointed out in
\citet{lhhbi01} \citep[also][]{grupe01}, there is no obvious reason
why intrinsically X-ray weak quasars should not exist.  Briefly, a
quasar arguably cannot exist without accretion as a source of fuel.
In an object like PHL~1811, that accretion probably occurs through an
optically thick, geometrically thin accretion disk that emits the
observed strong optical and UV continuum (Fig.\ 8). Such a disk would
never be hot enough  to emit X-rays, which are probably emitted by the
corona, a separate component.  We postulated that there may be
situations in which the corona may not exist or may be weak.

Why might the corona be weak?  As discussed in \citet{lhhbi01},
PHL~1811 is a very luminous NLS1.  As discussed by \citet{laor00},
narrow Balmer lines in luminous AGNs may imply high accretion rates if
the width of the lines is dominated by virial motions.  One possible 
central engine geometry considers the  X-rays to be emitted by a
central, hot, optically thin, geometrically thick disk, and the
optical and UV emitted by an optically thick, geometrically thin disk
with a large inner radius \citep[e.g.,][]{zg04}.  At high accretion
rates, it might be expected that the inner radius of the optically
thick, geometrically thin disk would shrink down toward the innermost
stable orbit, with the volume of the central hot X-ray emitting region
shrinking with it, and the intensity of the X-ray emission
correspondingly decreasing. This is thought to happen in Galactic
X-ray transient objects \citep[e.g.,][]{kb04}.   

Alternatively, the corona may lie on top of the optically thick,
geometrically thin accretion disk and be fed by reconnecting magnetic
flux tubes buoyantly emerging from the disk.  Decreasing the amount 
of energy released by reconnection in PHL~1811 would result in weak X-ray
emission.  Why would that happen?  One model, proposed by
\citet{bechtold03}, explains the dependence of $\alpha_{ox}$ on 
luminosity in the context of a disk/corona model in which the two
phases are thermally coupled, and in which the amount of energy
dissipated into the corona depends on the gas pressure in the disk.
In this model, $\alpha_{ox}$ is steeper for larger black holes, larger
accretion rates with respect to Eddington, and larger viscosity
parameters.  

It is also possible that since the optical-UV spectrum is very soft,
softer than most quasars, the corona is flooded by soft photons,
catastrophically Compton cooling it, and so reducing the X-ray
emission.  This idea is discussed  by \citet{proga05}, who suggests
that a luminous accretion disk can simultaneously drive an outflow and
quench the corona.  This idea is supported by the fact that we see
blueshifted high-ionization lines in PHL~1811 (discussed in Paper II),
and by the observed inverse correlation between $\alpha_{ox}$ and the
blueshift of \ion{C}{4} lines among NLS1s \citep{lm04}.  

Another way that a high accretion rate may lead to reduced X-ray
emission is through so-called ``photon trapping''.  If the accreting
gas is simply free-falling into the black hole, the photons may be
accreted before they can diffuse out through the accreting gas when
the accretion rate is high; they are trapped and advected into the
black hole \citep{begelman79}. Because the X-rays are most likely to
be emitted very close to the black hole, more of them could be trapped
than optical/UV photons, resulting in a steep $\alpha_{ox}$.  

\subsection{The Black Hole and  Accretion Disk in PHL 1811}

It is generally thought that the source of the optical-UV bump in AGN
broad band spectra is thermal emission from the accretion disk that
powers the active nucleus.  Despite this conviction, results from
observations of the optical-UV properties in AGN do not conform to our
expectations of accretion disks; specifically, the continuum does not
have the expected slope, the disk emission does not appear to be
polarized, and there is no prominent Lyman edge feature \citep{kb99}.
PHL~1811 has a prominent big blue bump (Fig.\ 8); in this section we
explore the accretion disk explanation for this feature.

The simplest model of an accretion disk is constructed by assuming
that half of accretion energy heats the disk, and the disk radiates
locally like a black body \citep[e.g.,][]{fkr92}. This model has the
advantage that it is easy to compute, but the disadvantage that is
not thought to be physically realistic.  As shown below, however,
we can obtain some useful results by quantitatively comparing this
model with our data. 

We compute the spectra emitted by the accretion disk between $R=3.1 \,
R_S$ and $1000 \, R_S$ in the frequency range $\log(\nu)=14$--$19$
(Hz), for a range of black hole masses between $10^7\, M_\odot$ and
$10^{11}\, M_\odot$, and a range of specific accretion rates, {\it
\.m}$=${\it \.M}$/${\it \.M}$_{Edd}$, between 0.1 and 10, noting that
this simple geometrically thin, optically thick model should break
down at the higher specific accretion rates.  We assume that the disk
is observed face on ($\cos i=0$).  To compare the accretion disk
spectra with our data, we identify regions of the merged optical and
UV spectrum that appear not to be dominated by emission lines, and
obtain the average flux in these bands\footnote{The bands we use are
1089--1102, 1317--1352, 1460--1482, 2229--2244, 3011--3038,
4014--4058, 4698--4768, and 5700--5750.}.  We then compute the sum of
the mean deviation of the log of $\nu L_\nu$ at the mean frequencies
for each band from each model.  The contours of the mean deviation are
shown in the left panel of Fig.\ 9.  We find that only a relatively
small range of $M_{BH}$ and {\it \.m} match the data.  The best
fitting values are $M_{BH}=2.2 \times 10^{9} M_\odot$ and {\it
\.m}$=0.9$.  The best fitting continuum spectrum is seen in the right
panel of Fig.\ 9.

What is the mass of the black hole in PHL 1811?  We can estimate the
mass using any number of the relationships between mass, H$\beta$
velocity width, and $\nu L_{\nu}$ at 5100\AA\/ currently available.
We use the one given by \citet{vp06}.  We measure $\lambda
L_\lambda(5100$\AA\/) to be $3.6 \times 10^{45} \rm \, erg\, s^{-1}$
using a distance of 936.4 Mpc.  We note that the flux from the merged
spectrum appears to be consistent with the broad-band photometry.  The
mean flux density over a 980\AA\/ band centered at observed 4400\AA\/
is $1.0 \times 10^{-14} \rm \, erg\, s^{-1}\,
cm^{-2}$\AA\/$^{-1}$. The photometry value, $B=14.4$, corresponds to
$1.15 \times 10^{-14} \rm \, erg\, s^{-1}\, cm^{-2}$\AA\/$^{-1}$, and
so is consistent within 15\%.  Given that the observations were not
contemporaneous, this is quite good agreement.  As will be discussed
in Paper II, we measure the width of H$\beta$ to be $1943\rm \, km\,
s^{-1}$.  Using \citet{vp06} Eq.\ 5, we find the black hole mass to be
$1.8 \times 10^{8}\, M_\odot$, a factor of 12 smaller than the
estimate based on the accretion disk model above.  Thus, the disk is
radiating at a rate significantly higher than the Eddington value to
attain the luminosity in the optical--UV band pass that we observe.
To illustrate this, we plot the spectrum for $M_{BH}=1.8 \times
10^8\rm \, M_\odot$ and {\it \.m}=1.0 on Fig.\ 9.  

It is worth pausing to note that there are several reasons why this
may not be an accurate estimate of the black hole mass in very
peculiar objects such as PHL~1811.  As is discussed in the
introduction of \citet{vp06} and references therein, there are a
number of assumptions and suppositions that go into the construction
of reverberation-based black hole masses. The primary assumption is
that the line widths are virial.  The evidence that this is true comes
from a very few Seyfert 1.5 galaxies with excellent multiwavelength
monitoring data \citep{pw00}.  Narrow-line Seyferts 1s like 
PHL~1811 are different from these objects in that their H$\beta$ lines
do not seem to vary, although variability has been observed in
low-luminosity NLS1s such as NGC~4051.  The other evidence that the
line widths are virial comes from the fact that in lower-luminosity
reverberation-mapped AGN, the black hole mass estimated from the line
widths and continuum luminosity is consistent with that estimated from
the host galaxy stellar velocity dispersion and the $M_{BH}$--$\sigma$
relationship.  PHL~1811 is much different than these objects as it has
a much higher luminosity.  The uncertainty in the black hole mass in
the reverberation-mapped objects is given to be a factor of $\sim 2.9$
based on the scatter around the $M_{bh}$-$\sigma$ relationship
\citep{onken04}.  However, the deviation from the
reverberation-mapping black hole mass in PHL~1811 may be larger for
several reasons.  First, it is possible that the broad line regions in
Narrow-line Seyfert 1s have a flattened configuration; this has been
previously suggested by \cite{md02}. A flatter distribution in NLS1s
would decrease the observed velocity width, and may also naturally
yield the observed lower equivalent widths as a consequence of a
smaller covering fraction.  This might imply a larger black hole mass
in PHL~1811 than estimated by the regressions based on
reverberation-mapped AGN.  On the other hand, PHL~1811 has a soft
spectral energy distribution. This means that it produces a weaker
ionizing continuum compared with other quasars with similar optical
luminosity. Inverting the argument originally made by \citet{wb98}
(applied to NLS1s that were inferred to have a stronger ionizing
continuum because of their prominent soft excess), this implies that
the H$\beta$ lines should be produced closer to the continuum source
than in other NLS1s, which would require the black hole mass to be
smaller.  The point is that extrapolating the relationships obtained
from the reverberation-mapped AGN may be particularly dangerous for
extreme objects such as PHL~1811. 

Regardless, assuming the black hole mass estimated above, we infer that
the disk must be radiating at a rate very much higher than the
Eddington value to attain the observed optical--UV luminosity.   We
note we are not the first to infer super-Eddington radiation in NLS1s;
this was also found by \cite{ck04} using a different approach.   They
estimate the black hole masses from the H$\beta$ line widths, and then
compare the observed bolometric luminosity to the predicted, whereas
we instead compare the luminosity and shape of the accretion disk to
the observed continuum.  

What is the bolometric luminosity of PHL~1811?  We integrate over the
inferred broad-band continuum to determine this.  For wavelengths
shorter than one micron, we use the spectral energy distribution
inferred from the observed optical, UV and X-ray data.  It is
comprised of piecewise power laws, suitable for use in {\it Cloudy},
and can be seen in Fig.\ 12 of Paper II.  Longward of one micron, we
use the \citet{elvis94} average continuum spectrum, since it is seen
to match the 2MASS and {\it IRAS} photometry rather well (Fig.\ 8).
The result is $3.7 \times 10^{46}\rm \, erg\, s^{-1}$.  It is 10.2
times $\lambda L_\lambda(5100)$, and so $L_{bol}/\lambda
L_\lambda(5100)$ is very close to 9, the bolometric correction factor
often used for AGN \citep[and assumed   in the mass-luminosity
  relationship in ][]{petersonetal04}.   

The Eddington luminosity for a $1.8 \times 10^{8}\rm\, M_\odot$ black
hole is $2.25 \times 10^{46} \rm \, erg\, s^{-1}$.  This implies that
PHL~1811, overall, is radiating at about 1.6 times the Eddington
luminosity.   But as the accretion disk fits show, the
optical-UV region is radiating at a very super-Eddington rate.

The sum-of-black bodies spectrum is not a physically realistic model
for an AGN accretion disk, although since it is the sum of black bodies,
it will be the overall {\it brightest} disk for a homogeneous,
isotropically emitting disk.  It does not take into account vertical
disk structure, Comptonization, or other effects that are expected to
influence the disk spectrum.  What do we expect to see from a
realistic accretion disk spectrum in comparison with the sum of black
bodies model?

\citet{sh93, sh95} considered the effect of vertical structure and
radiative transfer in the accretion disk, and applied their results to
AGN and black hole candidates.  They find that the emerging spectrum
is significantly harder than the sum of black bodies disk, due to
electron scattering and the vertical temperature gradient.  This means
that for a given accretion rate and black hole mass, their computed
disk spectrum has proportionally more soft X-rays than the
sum-of-black-bodies disk, but at the expense of the UV.  In order for
the flux of such a disk spectrum to match the UV flux from PHL~1811,
an even larger black hole would be necessary.  In fact, this should be
generically true for any disk model that radiates at less than the
Eddington rate \citep[e.g.,][]{ksm01}.

More recently, models for disks with super-Eddington emission have
been proposed by several authors  \citep[e.g.,][]{begelman02}.  In the
\citet{begelman02} model, the disk is inhomogeneous, and radiation
leaks out through optically thin channels.  We infer that PHL~1811 is
radiating at a super-Eddington rate in the optical and UV.
Perhaps this means that the spectrum of a super-Eddington disk should
be similar to that of a typical quasar longward of $\sim 1550$\AA\/,
based on Fig.\ 8 of Paper II.  One might expect the super-Eddington
emission to be stronger closer to the black hole, where the shorter
wavelength emission originates.  Perhaps this is the origin of the
unusual rising continuum shortward of $\sim 1400$\AA\/ shown in Fig.\
8 of Paper II.

\subsection{The Host Galaxy of PHL  1811} 

\citet{magorrian98} discovered that black hole masses are correlated
with the luminosity of the host bulge or elliptical galaxy.
Later it was found that a  tighter correlation between the velocity
dispersion of the bulge of a galaxy and the mass of the nuclear
black hole exists  \citep{fm00,gebhardt00}.  The host galaxies of
luminous active galaxies, i.e., quasars, are almost exclusively
ellipticals, and their basic properties are the same as elliptical
galaxies without quasars \citep[e.g.,][]{dunlopetal03}.  Furthermore,
they obey the relationship between black hole mass and spheroid mass
observed in nearby galaxies \citep[e.g.,][]{dunlop04}.   PHL~1811 has
a rather large black hole ($\sim$1.8$\times 10^8\rm \, M_\odot$; \S
5.2) and therefore should have a large elliptical galaxy as a host.  An
object that should in principle be quite similar is the quasar 3C~273.
Its redshift is 0.158 and its black hole mass is inferred to be $8.8
\times 10^8\, M_\odot$ \citep{petersonetal04}.  \citet{martel03}
present the ACS coronagraph data from this object.  They find a large
elliptical galaxy with an inner region about 17~kpc in diameter
($6.5^{\prime\prime}$), and extended diffuse emission out to
6--12$^{\prime\prime}$ from the active nucleus.  A de Vaucouleurs fit
to the V \& I profiles yield effective 
radii of $\sim 2.2^{\prime\prime}$ and $\sim 2.6^{\prime\prime}$.  

\citet{jenkins05} show an {\it HST} ACS WFP image of PHL~1811.
Observations of the host galaxies of nearby quasars are best done
using the coronograph; however, there was not sufficient time in that
program to make the coronagraph image with an accompanying PSF
calibration exposure.  They instead made a 520 second observation
directly using the F625W (Sloan $ r^*$) filter, and corrected for the
QSO PSF using a nearby star with colors similar to those of PHL~1811.
The resulting image was surprising; it does not show a large
elliptical host like 3C~273; rather there are structures that look
like spiral arms on either size of the QSO.  These features are not an 
artifact of the PSF subtraction as they can even be seen in the image
before subtraction, and for other reasons discussed by
\citet{jenkins05}.   Thus, PHL~1811 appears to have a spiral host, and
appears to lack a large bulge that might be expected given its
luminosity. 

The 3C~273 observations were 9--10 times longer than the PHL~1811;
perhaps the PHL~1811 image is not sufficiently well exposed to observe
the bulge?  We can estimate the expected properties of a typical host
galaxy,  by using our black hole mass estimate to estimate the
mass of the spheroid, then determining the expected observational
properties of an elliptical galaxy having a spheroid of that mass. 

In \S 5.2  we showed that the luminosity of the optical--UV spectrum is
consistent with a black hole of $2.2 \times 10^{9} \,\rm M_\odot$.
However, the black hole mass based on AGN reverberation mapping
results given by \citet{vp06}, we obtain a much smaller black hole
mass of $1.8 \times 10^{8} \,\rm M_\odot$.   Using the relationship
between the black hole mass and the mass of the spheroid \citep[$M_{BH}=0.0012
  M_{sph}$;][]{dunlop04} yields $M_{sph}= 1.5 \times 10^{11}\rm \, M_\odot$
for the estimated $M_{BH}=1.8 \times 10^{8}\rm \, M_\odot$.

It has  been found that the host galaxies of quasars are
indistinguishable from elliptical galaxies.  Therefore, we can use the
properties of elliptical galaxies to estimate the size and brightness
of the putative  host.  The mass of the spheroid is related to the
effective radius by $r_e=2.6 (M_{sph}/10^{11})^{3/5}\rm \, kpc$ for
$H_0=70 \rm \, km\, s^{-1}\, Mpc^{-1}$ \citep{loew99}. Thus, for
PHL~1811, we expect an effective radius of 3.3 kpc.  At the
distance of PHL~1811, one arc second corresponds to $3.2 \rm \, kpc$,
so $r_e=3.3\rm\, kpc$ corresponds to $1^{\prime\prime}$.  The image 
presented by \citet{jenkins05} is not reliable within about
$1.5^{\prime\prime}$ due to the correction for the QSO PSF, but should
be fine at $\sim 2 r_e$.  

It is well known that the surface brightness of elliptical galaxies is
correlated with the effective radius \citep[e.g.,][]{dd87}.  There is,
however, considerable scatter, although \citet{md02} argue that the
rms scatter for AGN hosts is only 0.18 dex.  We use the results from a
recent analysis of elliptical galaxies in the Sloan Digital Sky Survey
\citep{bernardi03}.  For the effective radius of 3.3 kpc, their Fig.\
10 shows that the mean surface brightness is $\sim 20 \rm \, mag \,
arcsec^{-2}$, and almost all objects lie within 19--21$\rm \, mag \,
arcsec^{-2}$ in the $r^*$ filter. We need to account for cosmological
dimming, which will lower the observed surface brightness by a factor
$(1+z)^4=2.02$.  Then, the mean surface brightness is $20.75 \rm \, mag \,
arcsec^{-2}$, with a conservative range of 21.75--19.75$\rm \, mag \,
arcsec^{-2}$.

Since the image is not reliable at $1\, r_e$ due to psf distortion, we
need to estimate the surface brightness at $2\, r_e$.  Assuming a
de~Vaucouleurs profile, we can use this average to find the expected
surface brightness at 2 $r_e$ \citep[e.g.,][]{binney}. We find then
the mean surface brightness to be $22.25 \rm \, mag \, arcsec^{-2}$,
with a range of 23.25--21.25$\rm \, mag \, arcsec^{-2}$.

Next, we use the exposure time calculator for the {\it HST} ACS to
determine the expected signal-to-noise ratio of diffuse emission
obtainable in a 520 second exposure using the F625W filter.  We use
the built-in spectrum of an elliptical galaxy, and include the effect
of Galactic extinction by $E(B-V)=0.046$.  We find that the mean
surface brightness of $22.25 \rm \, mag \, arcsec^{-2}$ yields a
signal-to-noise ratio in a 2$\times$2 pixel box of 5.4, with a range
of 2.3--11.7 for the surface brightness range.  Thus, if PHL~1811 had
a normal elliptical galaxy host, it may have been detected at $2\,
r_e=2^{\prime\prime}$ in the \citet{jenkins05} image, assuming it is
on the brighter side of the range of elliptical galaxies.

Finally, it has been found that NLS1s frequently have a much larger
host galaxy bulge than expected based on their black hole mass estimated
from their emission-line widths \citep[e.g.,][]{gm04,mg05}.  In
addition, \citet{ryan06} find from near-IR imaging of NLS1s that the
bulge is about an order of magnitude larger than would be inferred
using the $M_{BH}$ -- $\sigma$ relationshiop and the black hole masses
estimated from reverberation mapping.  If PHL~1811 had followed this
trend,  the host galaxy bulge should have been very easy to see.  

\section{Summary}

In this paper, we report the results and analysis of two {\it Chandra}
observations that were coordinated with an {\it HST} observation, an
{\it XMM-Newton} observation, two {\it Swift} ToOs, and MDM optical
photometry of the  unusually luminous, nearby narrow-line quasar
PHL~1811.  Here we summarize the primary results of the paper.

\begin{itemize}
\item The two 10ks {\it Chandra} observations, made twelve days apart,
  reveal a weak X-ray source at the position of PHL~1811.  A factor of
  four variability was observed between the two observations.  The
  X-ray spectrum is steep, with photon index $\Gamma \sim 2$--$2.6$,
  typical of the power law indices observed in {\it ASCA} spectra from
  NLS1s. There is marginal evidence of a soft excess in the brighter
  spectrum, and  marginal evidence for spectral variability in which
  the soft  excess became more prominent when the object was brighter.
  No evidence for absorption was found.

\item The $\sim 30$ ks {\it XMM-Newton} observation revealed no
  evidence for X-ray variability, and a flux consistent with the
  fainter of the two {\it Chandra} spectra.  The X-ray spectrum is
  steep with $\Gamma=2.3 \pm 0.1$, again typical of the power law
  indices observed from NLS1s by {\it ASCA}.   There is no evidence
  for absorption,   with an upper limit on intrinsic absorption of
  $8.7 \times 10^{20}\,   \rm cm^{-2}$.   The Optical Monitor data
  revealed no evidence for   short time-scale  UV variability. Two
  {\it Swift} observations found the flux to be consistent with the
  brighter of the   two {\it Chandra} spectra. 
  Overall, a factor of $\sim 5$ variation in X-ray flux was observed
  among the five X-ray observations presented   in this paper.  The UV 
  photometry, obtained by {\it XMM-Newton} and   {\it Swift}, is
  consistent with the {\it HST} spectrum discussed in   Paper~II; thus
  no evidence for UV variability has been found.  In addition, MDM
  photometry observations revealed no evidence for optical variability
  over three nights.

\item We compare the {\it XMM-Newton} PN spectrum of PHL~1811 with
  those from two representative NLS1s: Ton~S180, which has been
  previously modelled using a dual Comptonization model, and
  1H~0707$-$495, which has been previously modeled using partial
  covering or a reflection model.  Since the comparison
  NLS1s are much brighter than PHL~1811, we extracted spectra from
  sufficiently short segments that the number of photons in the PN
  spectra are approximately the same as in the PHL~1811
  spectrum. Fitted with a power-law model, the comparison spectra are
  significantly steeper than that of PHL~1811, and complexity is
  detected in the spectrum from 1H~0707$-$495.  PHL 1811's X-ray
  spectrum seems to be different from those of other NLS1s.  Since the 
  slope is consistent with the power-law observed in {\it ASCA}
  spectra of NLS1s, the simplest explanation for the X-ray spectrum in
  PHL~1811 is that it is powered by Compton upscattering of soft
  photons by energetic electrons, the typical power-law X-ray
  emission mechanism in quasars. 

\item The {\it BeppoSAX} observation found PHL~1811 to be significant
  X-ray weak \citep{lhhbi01}.  At that time, we presented three
  possible explanations for its X-ray weakness: 1.)  PHL~1811 is a
  BALQSO; 2.)  PHL~1811 is temporarily in a low flux state; 3.)
  PHL~1811 is intrinsically X-ray weak.  The spectral energy
  distribution, constructed using the {\it  Chandra} spectra and the
  {\it HST} spectrum taken two days before the first {\it Chandra}
  observation, confirms that PHL~1811 is X-ray weak, with
  $\alpha_{ox}$ between $-2.2$ and $-2.4$.  The {\it XMM-Newton} and
  {\it Swift} observations are consistent with these values.  The
  $\alpha_{ox}$ for a typical quasar of similar UV luminosity is
  $-1.6$, and compared with a sample of optically-selected quasars of
  similar UV luminosity, the X-ray flux of PHL~1811 is a factor
  13--450 times  weaker.  The lack of  absorption in the X-ray
  spectra, and the lack   of absorption lines in   the UV spectrum
  (Paper II) conclusively   rules out the possibility that PHL~1811 is
  a BALQSO, with a weak,   absorbed X-ray spectrum characteristic of
  that class of object.     PHL 1811 has now been
  observed seven times in X-rays between 1990 (during the {\it ROSAT}
  All Sky Survey) and the most recent {\it Swift} observation 2006 May
  14.  While it varies by about a factor of 5 among the latter 5
  observations, ruling out a scattering origin for the X-ray emission
  from an  extended electron-scattering mirror, it is always significantly
  weaker than other quasars of similar optical luminosity. While we 
  can never rule out the possibility that PHL~1811 is coincidentally
  observed always when it is in a transient  low state, the
  plausibility of this hypothesis is decreasing. We conclude that PHL
  1811 is {\it intrinsically} X-ray weak. 

\item We discuss the possible origins of the X-ray weakness.  It may
  be a consequence of the high accretion rate inferred from the narrow
  lines and high luminosity.  The corona may be smaller, and with a
  larger fraction of the accretion power going into the
  optically-thick, geometrically-thin accretion disk emitting the
  optical and UV continuum.  Alternatively, the corona may be quenched
  by the strong optical and UV emission, and may be unable to
  Compton-upscatter those photons to X-ray energies.  Another
  possibility is that high accretion rate traps the X-ray photons and
  advects them into the black hole.  

\item We compare the spectral energy distribution with a simple
  sum-of-black bodies accretion disk model.  We find that the UV
  spectrum is consistent with a black hole mass of $2.2 \times
  10^{9}\rm \, M_\odot$ and a specific accretion rate of {\it
  \.m}$=0.9$.  Using the H$\beta$ FWHM, the luminosity at 5100\AA\/
  and an equation derived from results of reverberation mapped AGN
  given by \citet{vp06}, we obtain a black hole mass estimate of $1.8
  \times   10^8 \rm \, M_\odot$, a factor of 12 below the value
  obtained from the   accretion disk spectrum.  We caution that we do
  not have a direct measurement of the black hole mass for this
  object. We point out that while the sum-of-blackbodies disk is not
  physically realistic, it is the brightest homogeneous,
  isotropically-emitting disk, and most other,  more realistic, disks
  will require an even higher radiation rate relative to Eddington.
  The exception may be inhomogeneous disks recently proposed by e.g., 
  \citet{begelman02} in which radiation leaks out through
  optically-thin channels.   Based on the observed
  broad-band continuum   spectrum, and extrapolating to the IR using
  the \citet{elvis94} SED,   we estimate the bolometric luminosity 
  to   be $3.7 \times   10^{46}\rm erg\, s^{-1}$.  Thus, overall, 
  PHL~1811 is    radiating at 1.6 times the Eddington rate, with the
  optical and UV   well above Eddington.    We  also discuss the {\it
  HST} ACS image presented by \citet{jenkins05} that reveals a spiral
  galaxy at the position of PHL~1811.  Most quasars have elliptical
  hosts; we   demonstrate that if PHL~1811 had the large elliptical
  host typical   of average quasars with $1.8 \times 10^8\rm\,M_{BH}$
  black holes, it   might have been seen (i.e., the estimated SNR is
  $\sim 5$) in that   image.  A spiral host galaxy appears to be yet
  another unusual  feature of PHL 1811.    

\end{itemize}

\newpage



\acknowledgments
KML acknowledges helpful conversations with
Andrzej Zdziarski regarding accretion disks, and Mike Loewenstein
regarding elliptical galaxies. We thank the anonymous referee for
helpful comments.  Support for this work was provided by
the National Aeronautics and Space Administration through Chandra
Award number GO2-3119X issued by the Chandra X-ray Observatory Center,
which is operated by the Smithsonian Astrophysical Observatory for and
on behalf of the National Aeronautics Space Administration under
contract NAS8-39073. KML and JC gratefully acknowledge support
through grant NNG05GB38G ({\it XMM-Newton} Cycle 3 Guest Observer).
This research has made use of the NASA/IPAC Extragalactic Database
which is operated by the Jet Propulsion Laboratory, California
Institute of Technology, under contract with the National Aeronautics
and Space Administration.




\clearpage

\begin{deluxetable}{lcccc}
\tablewidth{0pt}
\tabletypesize{\scriptsize}
\tablecaption{Observing log}
\tablehead{
\colhead{Observatory andr} & \colhead{Date} &  
 \colhead{Exposure} & \colhead{Bandpass or} &
 \colhead{Extraction Radius} \\
& \colhead{Instrument} & \colhead{(seconds)} & \colhead{Effective Wavelength} & \colhead{(arcseconds)}}
\startdata

{\it Chandra} ACIS S3 & 2001 Dec 5 &  9377 & 0.3--10 keV & $3.94$ \\
{\it Chandra} ACIS S3 & 2001 Dec 17  &  9839 & 0.3--10 keV & $3.94$ \\
{\it XMM-Newton} PN & 2004 Nov 01 & 27472 & 0.3--12 keV  & $27$ \\
{\it XMM-Newton} MOS1 & \ldots & 32126 & 0.5--10 keV & $23$\\
{\it XMM-Newton} MOS2 & \ldots  & 32164 & 0.5--10 keV & $23$ \\
{\it XMM-Newton} OM (UVM2)  & \ldots & 25800 & 2310\AA\/ & 12 \\
{\it Swift} XRT (PC mode) & 2005 Oct 22 & 2462 & 0.3--10 keV & $23.4$\\
{\it Swift} UVOT (UVW1) & \ldots & 787 & 2600\AA\/ &  12  \\
{\it Swift} UVOT (UVM2) & \ldots & 844 & 2200\AA\/ & 12 \\
{\it Swift} UVOT (UVW2) & \ldots & 844 & 1930\AA\/ &  12 \\
{\it Swift} XRT (PC mode) & 2006 May 12  & 1600  & 0.3--10 keV & $23.4$ \\
{\it Swift} UVOT (V) & \ldots & 130 & 5460\AA\/ & 6 \\
{\it Swift} UVOT (B) & \ldots & 130 & 4350\AA\/ & 6 \\
{\it Swift} UVOT (U) &  \ldots & 130 & 3450\AA\/ & 6 \\
{\it Swift} UVOT (UVW1) & \ldots & 259 & 2600\AA\/ & 12 \\
{\it Swift} UVOT (UVM2) & \ldots & 372 & 2200\AA\/ &  12  \\
{\it Swift} UVOT (UVW2) & \ldots & 526 & 1930\AA\/ & 12  \\
MDM McGraw-Hill 1.3 + I + Templeton & 2004 Oct 14--16 & 5,3,1\tablenotemark{a} & 8050\AA\/ & 5.8--10.0 \\
MDM McGraw-Hill 1.3 + V + Templeton & \ldots & 5,9,5\tablenotemark{a} & 5380\AA\/ & 5.7--10.1 \\
MDM McGraw-Hill 1.3 + B + Templeton & \ldots & 5,5,2\tablenotemark{a} & 4350\AA\/ & 6.2--10.0 \\
MDM McGraw-Hill 1.3 + U + Templeton & \ldots & 5,5,2\tablenotemark{a} & 3640\AA\/ & 6.8--10.2 \\
\enddata
\tablenotetext{a}{Number of frames on each of the three days.}
\end{deluxetable}
\normalsize

\begin{deluxetable}{lcccc}
\tablewidth{0pt}
\tablecaption{{\it Chandra} Spectral Fitting Results}
\tablehead{
\colhead{Parameter} & \colhead{2001--12--05} & \colhead{2001--12--17}}
\startdata

\sidehead{1. Power law model, spectra fit separately.}

Photon Index & $2.01^{+0.37}_{-0.36}$ & $2.58^{+0.19}_{-0.18}$ \\
Normalization\tablenotemark{a} & $1.17\pm 0.02 \times 10^{-5}$ &
$5.10 \pm 0.45 \times 10^{-5}$ \\
Flux (0.3--5 keV; $\rm erg\,cm^{-2}\,s^{-1}$) & $5.2 \times 10^{-14}$
& $2.3 \times 10^{-13}$ \\
Luminosity (0.3--5 keV; $\rm erg\, s^{-1}$) & $5.6 \times 10^{42}$ &
$2.4 \times 10^{43}$ \\
$\chi^2$/d.o.f & 1.96/2 & 13.2/15 \\
\tableline
\sidehead{2. Power law + blackbody\tablenotemark{b}}
Photon Index &\nodata & $2.22 \pm 0.34$ \\
PL Normalization\tablenotemark{a} & \nodata & $4.4^{+0.83}_{-1.0} \times 10^{-5}$ \\
Blackbody temperature (keV) & \nodata & $0.096^{+0.040}_{-0.081}$ \\
BB Normalization\tablenotemark{c} & \nodata & $1.5_{-1.1}^{+\infty}\times 10^{-6}$ \\
$\chi^2$/d.o.f & \nodata & 8.36/13 \\
\tableline
\sidehead{3. Power law + blackbody\tablenotemark{b}, joint fit}
Photon Index & $2.12 \pm 0.25$\tablenotemark{d} &  $2.12 \pm 0.25$\tablenotemark{d}  \\
PL Normalization\tablenotemark{a} & $1.17\pm 0.22 \times 10^{-5}$ &
$4.14^{+0.79}_{-0.84} \times 10^{-5}$ \\
Blackbody temperature (keV) &
$0.10^{+0.035}_{-0.042}$\tablenotemark{d} &
$0.10^{+0.035}_{-0.042}$\tablenotemark{d} \\ 
BB Normalization\tablenotemark{c} & 0\tablenotemark{e} &
$1.5^{+2.7}_{-0.82} \times 10^{-6}$ \\
$\chi^2$/d.o.f & 10.8/16\tablenotemark{d} & 10.8/16\tablenotemark{d} \\
\enddata
\tablenotetext{a}{In units of $\rm photons\,
  keV^{-1}\,cm^{-2}\,s^{-1}$ at 1 keV in the observed frame.}
\tablenotetext{b}{The blackbody temperature is given in the  rest frame.}
\tablenotetext{c}{In units of $L_{39}/D_{10}^2$, where $L_{39}$ is the
  source luminosity in units of $10^{39}\rm\, erg\, s^{-1}$, and
  $D_{10}$ is the distance to the object in units of 10 kpc.}
\tablenotetext{d}{Parameter fit jointly.}
\tablenotetext{e}{Fixed parameter.}

\end{deluxetable}

\clearpage

\begin{deluxetable}{lc}
\tablewidth{0pt}
\tablecaption{{\it XMM-Newton} Spectral Fitting Results}
\tablehead{
\colhead{Parameter} & \colhead{Measurement} }
\startdata

Photon Index & $2.28^{+0.12}_{-0.11}$  \\
PN Normalization\tablenotemark{a,b} & $1.16 \pm 0.98$ \\
MOS1 Normalization Offset\tablenotemark{a} & $1.25^{+0.21}_{-0.19}$ \\
MOS2 Normalization Offset\tablenotemark{a} & $1.32^{+0.22}_{-0.20}$ \\
Flux (0.3--5 keV; $\rm erg\, cm^{-2}\,s^{-1}$) & $5.1 \times 10^{-14}$ \\
Luminosity (0.3--5 keV; $\rm erg\,s^{-1}$) & $5.4 \times 10^{42}$ \\
$\chi^2$/d.o.f. & 74.7/79 \\
\enddata
\tablenotetext{a}{The spectral model was $C e^{N_H(Gal) \sigma(E)}
  E^{-\Gamma}$, where all parameters were tied together except for the
  constant $C$, which was fixed to 1 for the PN and allowed to vary
  for the MOS1 and MOS2.  Unexpected optical loading plausibly causes
  the MOS spectra to   have a higher normalization; therefore we quote
  flux and luminosity values from the PN data only.}
\tablenotetext{b}{In units of $\rm photons\,
  keV^{-1}\,cm^{-2}\,s^{-1}$ at 1 keV in the observed frame.}
\end{deluxetable}

\clearpage

\begin{deluxetable}{lcccccc}
\tablewidth{0pt}
\tablecaption{{\it Swift} UVOT Results}
\tablehead{
 & \multicolumn{6}{c}{Flux Density ($10^{-14}\rm
  erg\,s^{-1}\, cm^{-2}\, $\AA\/$^{-1}$)\tablenotemark{a}} \\
\colhead{Observation} & \colhead{UVW2} & \colhead{UVM2} & \colhead{UVW1} & \colhead{U} &
\colhead{B}  & \colhead{V}}

\startdata

2005 October 22 & $3.75 \pm 0.03$ & $2.75 \pm 0.02$ & $2.45 \pm 0.02$
& \nodata & \nodata & \nodata \\
2006 May 12 & $3.66 \pm 0.03$ & $2.71 \pm 0.03$ & $2.47 \pm 0.03$ &
$1.89 \pm 0.04$ & $0.99 \pm 0.02$ & $0.61 \pm 0.02$ \\
\enddata
\tablenotetext{a}{Observed fluxes; uncorrected for Galactic reddening.}
\end{deluxetable}

\clearpage


\begin{figure}
\plotone{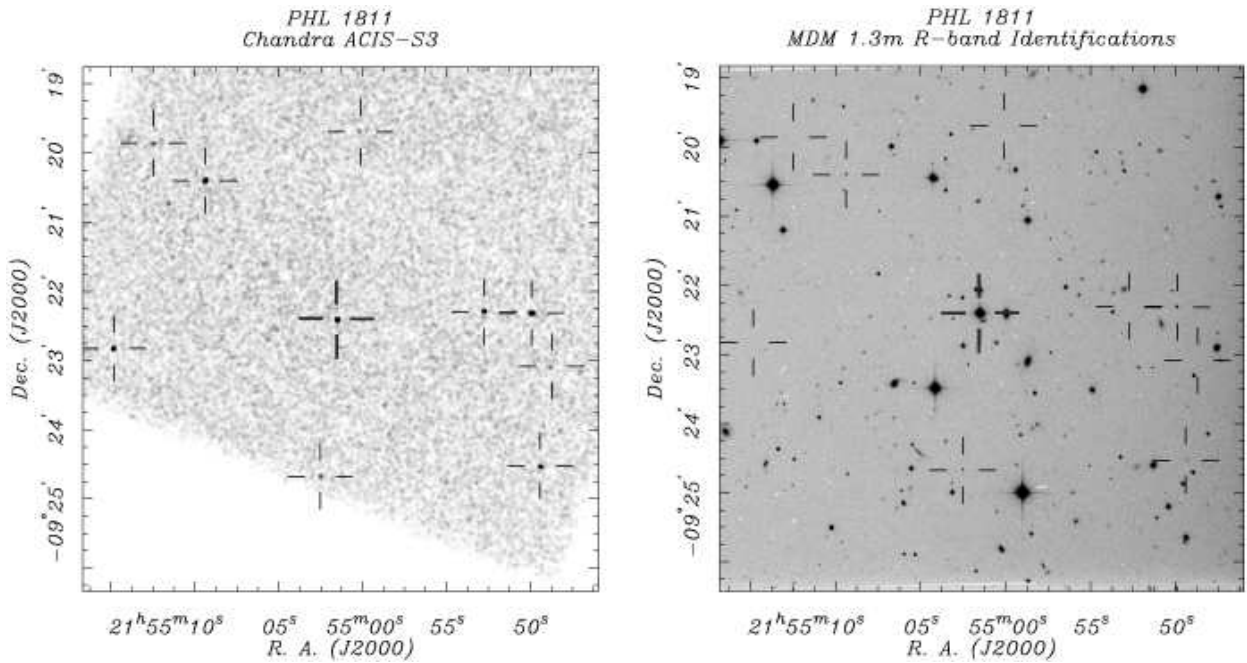}
\caption{The {\it Chandra} image ({\it left}) and the MDM optical {\it
  R}-band   image ({\it right}) show that the X-ray source is securely
  identified as PHL 1811. 
\label{fig1}} 
\end{figure}

\begin{figure}
\plotone{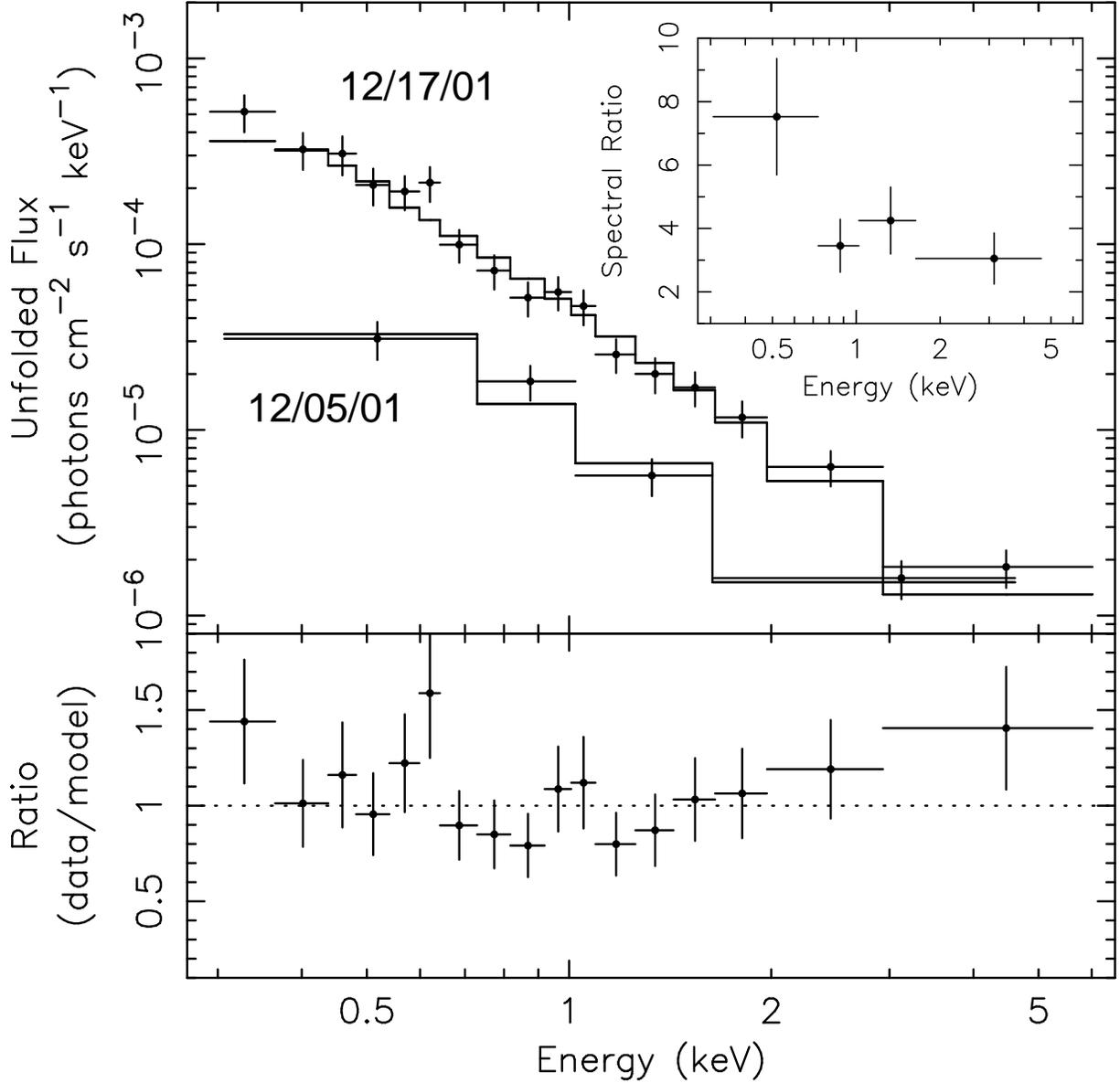}
\caption{{\it Top}: Unfolded best fitting power-law models for the spectra
  from the two {\it Chandra} observations.  {\it Bottom}: Residuals from a
  power-law model   fit to the spectrum from the second observation
  have a convex shape   suggesting the presence of a soft
  excess. {\it Inset:} The ratio of second   spectrum to the first spectrum
  shows that the spectral variability   is predominately confined to
  the softest energies, suggesting the   emergence of a soft excess
  component when the object is bright. 
\label{fig2}} 
\end{figure}

\begin{figure}
\epsscale{1.0} \plotone{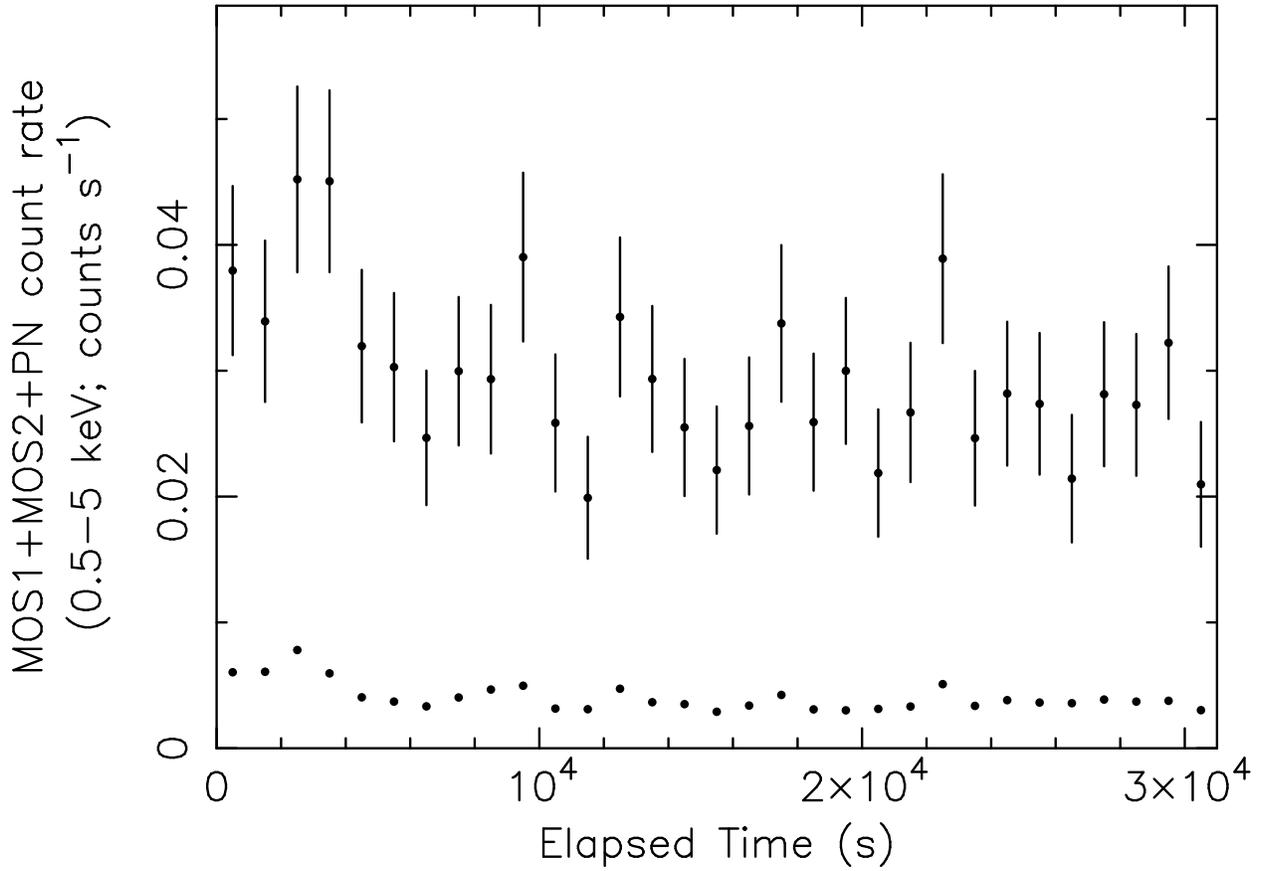}
\caption{{\it XMM-Newton} background-subtracted lightcurve composed of
  the 0.5--5 keV photons from the MOS1, MOS2 and PN detectors.  The
  lower light curve shows the estimated background in the source
  extraction region. There is no strong   evidence for variability
  over the 31~ks   observation. 
\label{fig3}} 
\end{figure}

\begin{figure}
\epsscale{1.0} \plotone{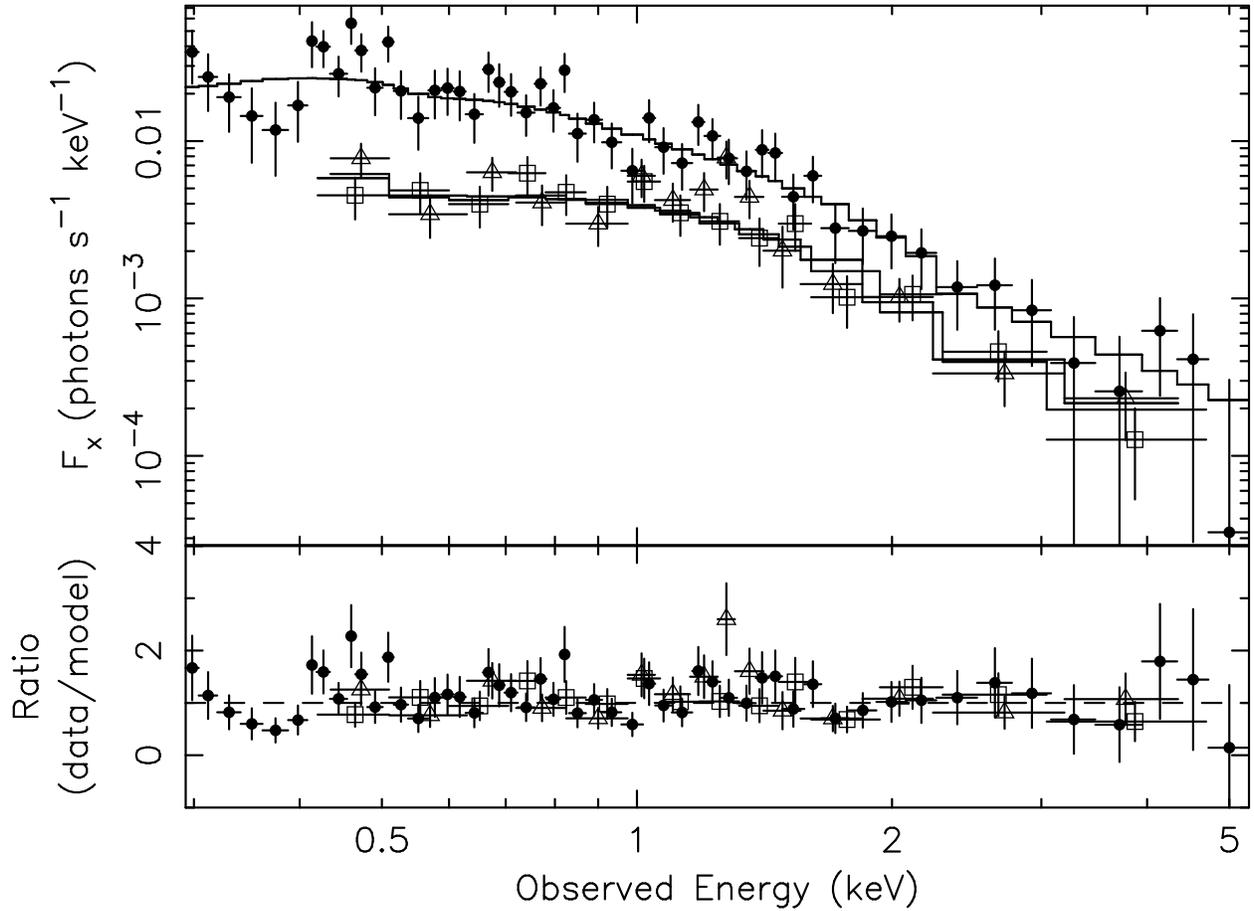}
\caption{The {\it XMM-Newton} MOS1, MOS2 and PN spectra (open squares,
  open triangles, and filled circles, respectively) are well fit
  using a power law and Galactic absorption.  The upper limit on
  intrinsic X-ray absorption is $8.7 \times 10^{20} \rm cm^{-2}$.
\label{fig4}} 
\end{figure}

\begin{figure}
\epsscale{1.0} \plotone{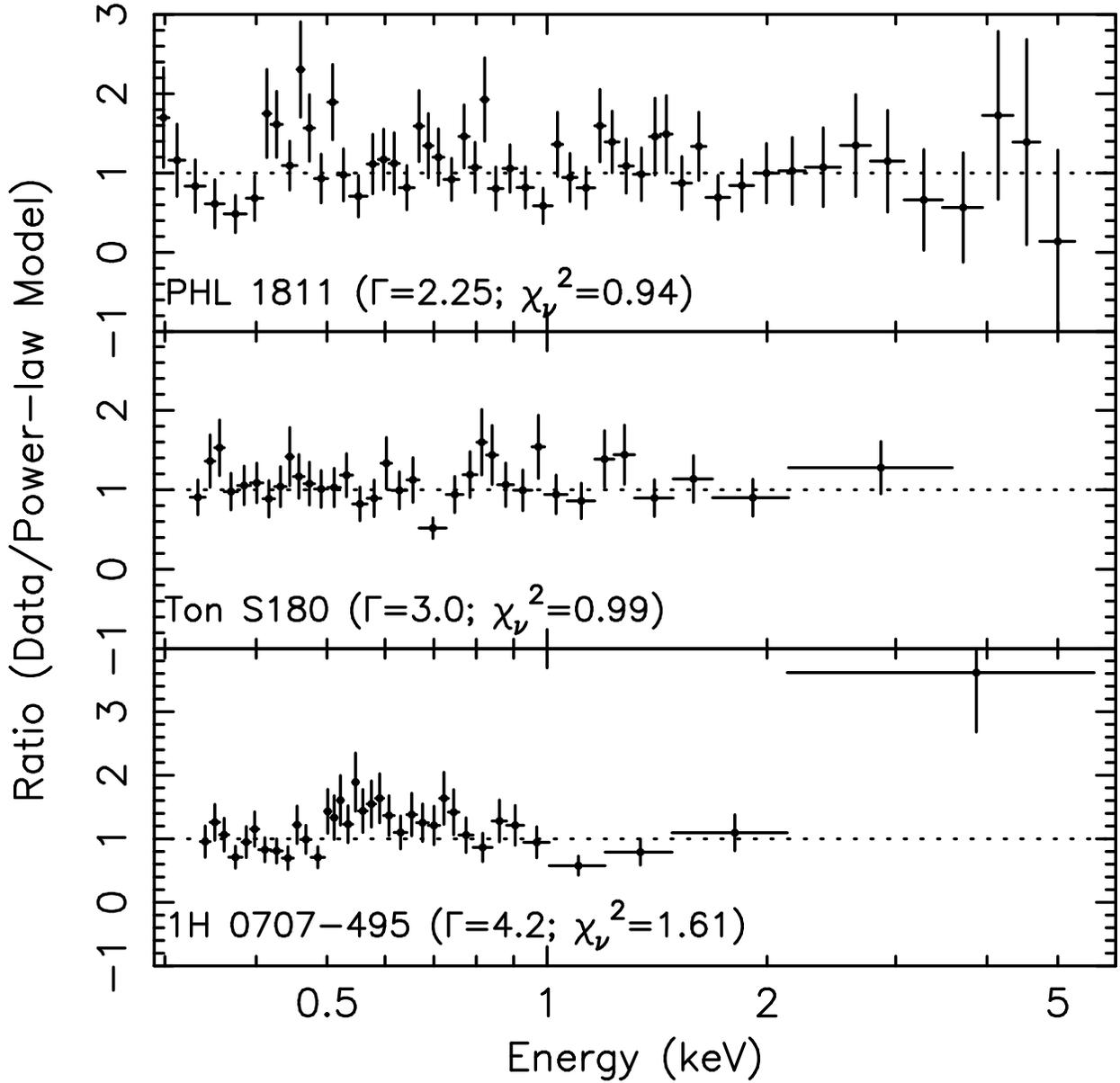}
\caption{A comparison of the {\it XMM-Newton} PN spectrum from
  PHL~1811 with spectra from other NLS1s from segments of data short
  enough for the statistics to be similar.  In each panel, the ratio
  of the data to a power law plus Galactic absorption model is
  plotted.  The resulting power law index and $\chi_\nu^2$ is given in
  each panel.  {\it Top:} PHL~1811 is statistically well described by
  a power law model.  The photon index is typical of the power law
  indices observed from  {\it ASCA} observations of NLS1s
  \citep{leighly99b}.  {\it Middle:} Ton~S180, an 
  NLS1 classified as having a ``simple'' spectrum by \citet{gallo06}, is
  statistically well described by a power law; however, the index is
  steep, suggesting that a longer exposure would reveal a hard tail
  \citep[as it does,][]{vaughan02}.  {\it Bottom:} 1H~0707$-$495, an
  NLS1 classified as having a ``complex'' spectrum by \citet{gallo06},
  shows a very steep spectrum and significant residuals, confirming
  the necessity of a complex model \citep[e.g.,][]{gallo04b,
  tanaka04}. 
\label{fig5}} 
\end{figure}

\begin{figure}
\epsscale{1.0} \plotone{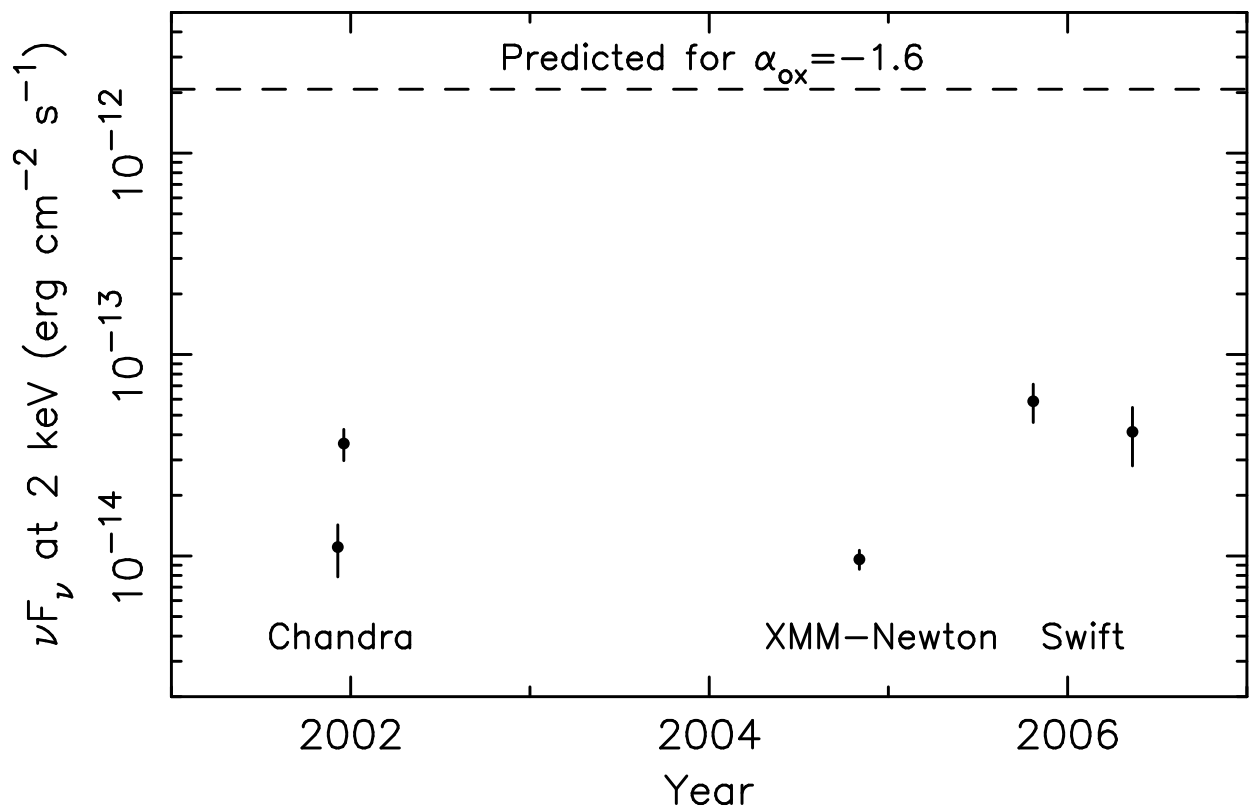}
\caption{Long-term X-ray flux variability of PHL 1811 in terms of $\nu
  F_{\nu}$ 
at 2 keV in the rest frame.  The uncertainties for the spectroscopic
data ({\it Chandra} and {\it XMM-Newton}) are propagated 1-$\sigma$
errors in the power law normalization and index.  For the detection
data ({\it Swift}) the uncertainties are proportional to the count
rate error. Note the logarithmic flux axis.  Also shown is the
predicted flux for $\alpha_{ox}=-1.6$ based on the UV flux from the
{\it HST} observations.
\label{fig6}} 
\end{figure}

\begin{figure}
\epsscale{1.0} \plotone{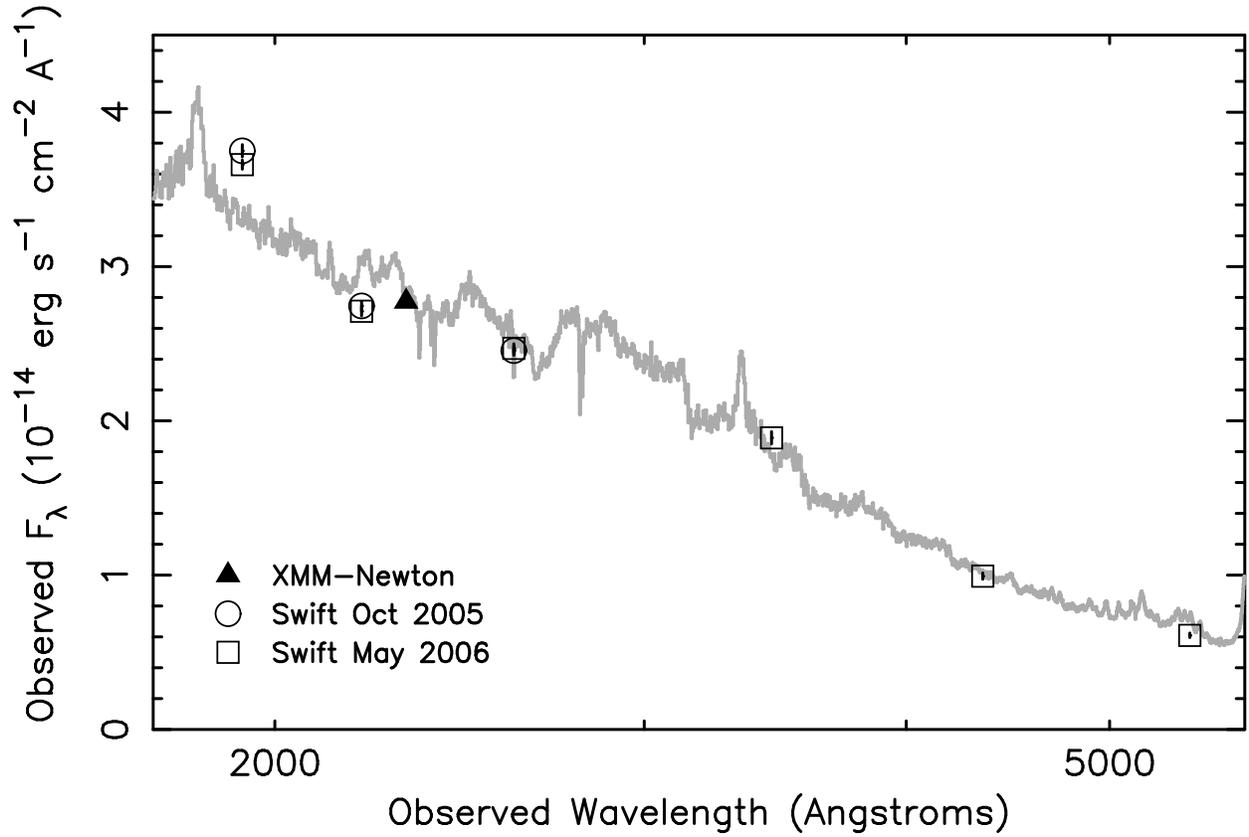}
\caption{The {\it XMM-Newton} OM and {\it Swift} UVOT photometry
  points are overlaid on the observed merged optical and {\it HST} UV
  spectrum observed 2001 Dec 03 and discussed in Paper II.  No strong
  evidence for UV variability is seen among the observations which
  span 4.5 years. 
\label{fig7}} 
\end{figure}

\begin{figure}
\epsscale{1.0} \plotone{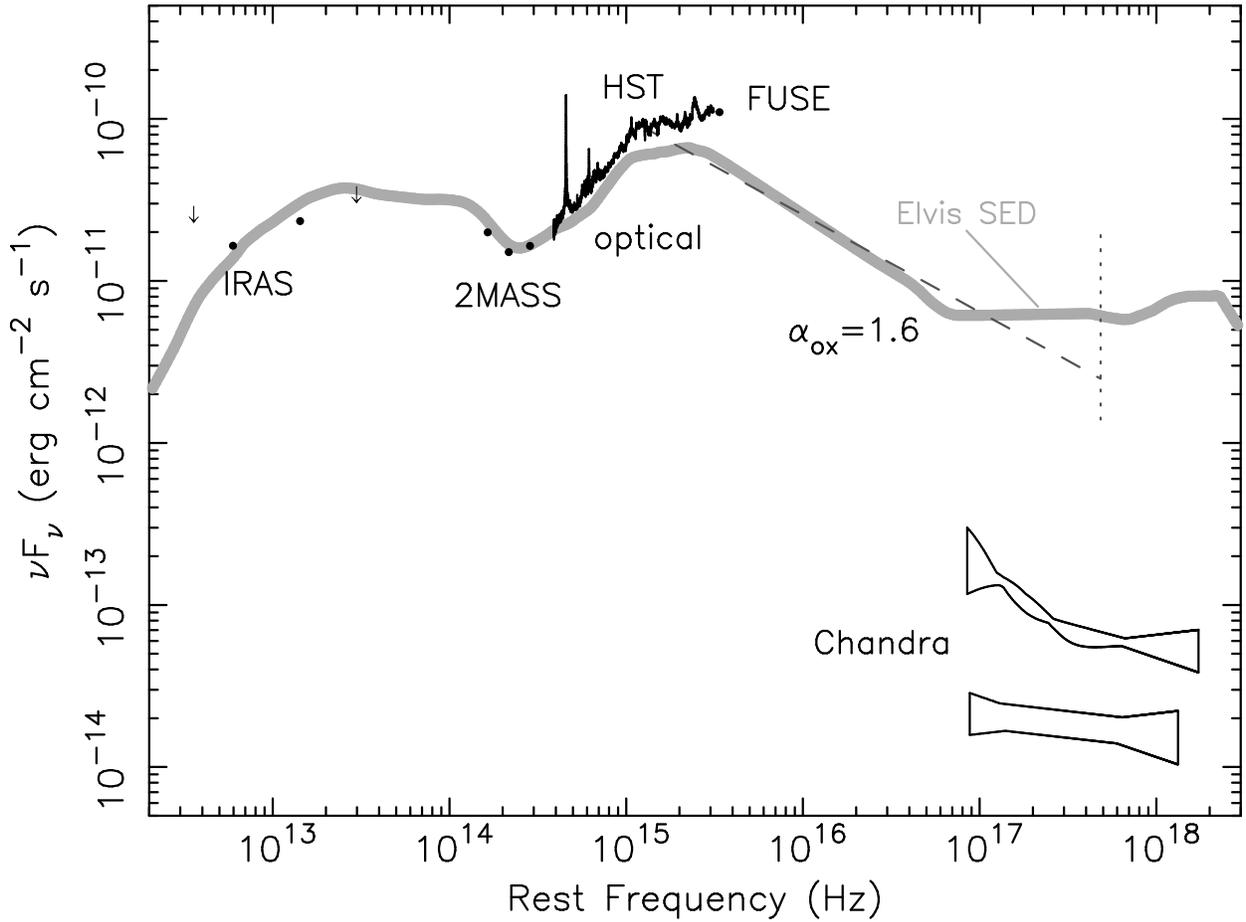}
\caption{The spectral energy distribution of PHL 1811, plotted as a
  function of the rest-frame frequency.  Contours from each of the two
  {\it Chandra} observations are shown, generated by successively
  setting each parameter to its $\Delta \chi^2=2.71$ value and
  computing the model, then determining the maximum and minimum of all
  of the models.  The dashed line shows the
  expected 2~keV flux for an average quasar of this luminosity, based
  on the regression presented by \citet{wilkes94}, while  the dotted
  line shows the range observed by  \citet{wilkes94}.  The average
  quasar SED from \citet{elvis94}, scaled to the 1 micron inflection,
  is also shown.
\label{fig8}} 
\end{figure}

\begin{figure}
\epsscale{1.0} \plotone{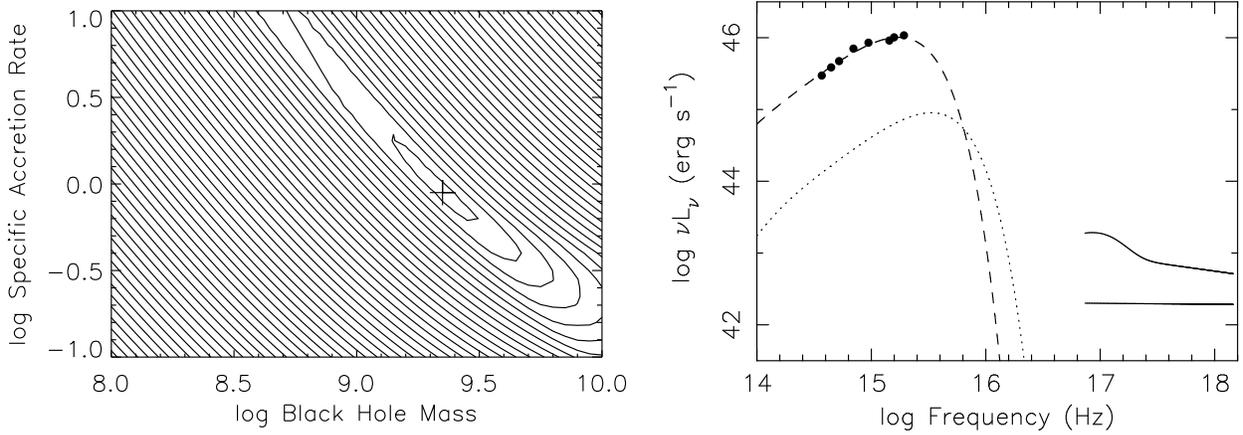}
\caption{{\it Left:} Contours of equal deviation of UV flux points
  from sum-of-black bodies accretion-disk models, as a function of
  input black hole mass (in units of solar masses), and the specific
  accretion rate, {\it \.m}. See text for further information.  The contour
  interval is $0.05$ in units of $\log\nu L_\nu$.  The minimum is located at
  $M_{BH}=2.2 \times 10^9 \,   M_\odot$ and {\it \.m}=0.9.  {\it
  Right:} The UV continuum flux points and   the X-ray spectra are
  shown as solid dots and solid lines   respectively. The best fitting
  model is shown by the dashed line. For comparison, the spectrum for
  a $M_{BH}=1.8 \times 10^8 \,   M_\odot$ black hole with {\it
  \.m}=1.0 is shown by the dotted line.  
\label{fig9}} 
\end{figure}


\end{document}